\documentclass[onecolumn,superscriptaddress,nofootinbib,preprintnumbers]{revtex4-2}
\usepackage[colorlinks=true,breaklinks=true]{hyperref}
\usepackage[utf8]{inputenc}
\hypersetup{urlcolor=Blue,
	    citecolor=Blue,
	    linkcolor=Blue}
\usepackage{mathtools}
\usepackage[dvipsnames]{xcolor}
\usepackage{bbm}
\usepackage{orcidlink}

\usepackage{amsmath}
\usepackage{amssymb}

\begin{document}

\title{Heating the dark matter halo with dark radiation from supernovae}

\author{Stefan Vogl \orcidlink{0000-0002-3005-9279}}
\email{stefan.vogl@physik.uni-freiburg.de}
\affiliation{Institute of Physics, University of Freiburg\\Hermann-Herder-Str.~3, 79104 Freiburg, Germany}

\author{Xun-Jie Xu \orcidlink{0000-0003-3181-1386}}
\email{xuxj@ihep.ac.cn}
\affiliation{Institute of High Energy Physics, Chinese Academy of Sciences\\ Beijing 100049, China}
\affiliation{Peng Huanwu Center for Fundamental Theory\\ Hefei, Anhui 230026, China}
\preprint{USTC-ICTS/PCFT-24-52}
\date{\today}

\begin{abstract}

Supernova explosions are among the most extreme events in the Universe, making them a promising environment in which to search for the effects of light, weakly coupled new particles. As significant sources of energy, they are known to have an important effect on the dynamics of ordinary matter in their host galaxies but their potential impact on the dark matter (DM) halo remains less explored. 
In this work, we investigate the possibility that some fraction of the supernova energy is released via the form of dark radiation into the DM halo.
Based on evaluation of energetics, 
we find that even a small fraction of the total SN energy is sufficient to change the overall shape of the DM halo and transform a cuspy halo into a cored one. This may help to explain the cores that are observed in some dwarf galaxies. 
Alternatively, one can interpret the upper limit on the size of a possible DM core as an upper limit on the energy that can go into light particles beyond the SM. These arguments are largely independent of a concrete model for the new physics. Nevertheless, it is important to ensure that the conditions we need, i.e.~significant supernova emissivity of dark radiation  and the opacity of DM halo to the dark radiation,
can be met in actual models.
To demonstrate this, we study four simple benchmark models: the dark photon, dark Higgs, and gauged $B-L$ and $L_\mu - L_\tau$ models\,---\,all provide light weakly coupled particles serving as the dark radiation. 
Assuming a sizable coupling of the dark radiation to DM, we find that all of the benchmark models have a significant part of the parameter space that meets the conditions. Interestingly, the couplings allowed by observations of SN1987A 
can have a significant effect on the halo of dwarf spheroidal galaxies. 

\end{abstract}
 
\maketitle

\section{Introduction}
\noindent
Type-II supernova (SN) explosions are frequent and release a large amount of energy of  $\approx 3 \times 10^{53}$ erg~\cite{Raffelt:1990yz,Caputo:2024oqc}. 
In standard astrophysics, approximately $99\%$ of this is emitted in the form of neutrinos while only about $1\%$ goes into the spectacular explosion that can be observed over many wavelengths in optical channels. Due to the extreme conditions in the SN core, where the temperature can reach tens of MeV and the density goes up to $\gtrsim 10^{14} \mbox{g}/\mbox{cm}^3$ (see e.g.~\cite{2007PhR...442...38J} for a review), SN explosions offer unique conditions to test new physics. 
As the outer layers shield the inner part of the SN very efficiently, standard astrophysical observations are largely insensitive to the conditions in the core and it is difficult to make definitive statements about new physics. Therefore, most work in this direction has focused on the insights that can be gained from the neutrino emission of SN 1987A. The energy and the timing of the neutrino events are consistent with expectations from theoretical modeling of the explosion and match the comparatively slow cooling of a protoneutron star. 
Frequently, limits on new physics are based on the ``Raffelt criterion" \cite{Raffelt:1996wa}, i.e.~the average energy loss rate to these new states at times of $\approx 1$ sec has to be smaller than the total energy loss rate to neutrinos such that the duration of the cooling phase is not cut short. There is a large body of literature that investigates the implications of the observed neutrino signal on models with light new physics such as 
sterile neutrinos \cite{Raffelt:1987yt,Nunokawa:1997ct,Raffelt:2011nc,Arguelles:2016uwb,Brdar:2023tmi,Carenza:2023old}, 
axions and axion-like particles \cite{Turner:1987by,Chang:2018rso,Carenza:2019pxu,Springmann:2024mjp,Springmann:2024ret,Caputo:2024oqc,Carenza:2024ehj}, 
dark gauge bosons (including dark photon) \cite{Dent:2012mx,Kazanas:2014mca,Chang:2016ntp,Sung:2019xie, Croon:2020lrf,Hong:2020bxo,Shin:2021bvz,Shin:2022ulh}, 
dark Higgs \cite{Rrapaj:2015wgs,Krnjaic:2015mbs,Hardy:2016kme,Hardy:2024gwy}, 
and various other light bosonic or fermionic species with couplings to the constituents of the SN core \cite{Choi:1989hi,Heurtier:2016otg,Chu:2019rok,Bollig:2020xdr,Dev:2020eam,Camalich:2020wac,Caputo:2021rux,Fiorillo:2022cdq,Akita:2022etk,Akita:2023iwq,Asai:2022zxw,Fiorillo:2023cas,Manzari:2023gkt,Fiorillo:2023ytr,Lazar:2024ovc,Telalovic:2024cot,Fiorillo:2024upk}. 
Currently, further progress in this direction is hampered by the limited data collected from SN1987A. While theoretical improvements are still possible and highly desirable, a qualitative step forward will require the observation of the next galactic SN. Therefore, it is interesting to ask if there are other observables that may be sensitive to new physics in SN explosion.

It is known that the energy released from SN can have an effect on astrophysical observables. For example,
N-body simulations that include models for baryonic feedback find SN explosions could explain the formation of cores in DM halos. DM-only simulations predict halos to be cuspy \cite{Navarro:1996gj} but observations of some dwarf spheroidal galaxies (dSph) prefer cored halos, sees e.g.~\cite{Bullock:2017xww} for a review of the small scale problems of $\Lambda$CDM. 
From considerations of energetics  
it can be shown that SN release sufficient energy in the visible sector to enable a transformation of an initial cuspy DM halo to a cored one if a significant fraction of the energy can be transferred to the DM \cite{Penarrubia:2012bb}.
Recently \cite{Heston:2024ljf} has considered the impact of this energy release on the DM halo in a model where DM-neutrino scattering allows for an efficient transfer of the energy to DM halo. Similar arguments can be used to place limits on the amount of energy emitted in the form of very weakly coupled BSM particles such as the dark photon or a dark Higgs.   If these particles are able to transfer their energy to the halo they will also affect the shape of the halo. This allows us to test values of the coupling to the SM  that are orders of magnitude smaller than those excluded by the usual SN cooling arguments mentioned above. 
This is the main question we want to address in this paper.

This paper is organized as follows: First, we will introduce cored and cuspy profiles for the DM halo and infer the energy that is required to affect a cusp-core transformation from the difference of the gravitational binding energy. 
Combined with measurements of the halo mass and the density in the center from \cite{Read:2018fxs} this allows us to determine the amount of energy that can be injected into the halo. 
By comparing this with the total energy released from type-II SN explosions, we derive an upper limit on the fraction of energy that can be released via exotic cooling channels, provided that the energy is absorbed by the halo. 
These arguments are quite general and no reference to a concrete particle physics model is needed at this stage. In Sec.~\ref{sec:particle}, we move to particle physics and discuss the production of light bosons in the SN core using a $Z'$ with generic couplings as a template model. 
In the second part of the section, we investigate the constraints that an efficient transfer of energy to the DM halo places on the parameters of the model. 
Finally, in Sec.~\ref{sec:models} we study four representative benchmark models and confront the parameter space that allows for a large energy injection in the halo with other observations.
We present our conclusions in Sec.~\ref{sec:conclusion}. Technical details regarding some aspects of particle production in the SN core are provided in the Appendix.


\section{Supernova-induced cusp-core transformations \label{sec:framework}}
\subsection{Halo profiles and gravitational binding energy}
This discussion largely follows arguments first presented in \cite{Penarrubia:2012bb} for the effect of baryonic feedback on the DM halo.
From DM only N-body simulations, the density of the halo is expected to follow an NFW profile \cite{Navarro:1996gj}:
\begin{align}
\label{eq:NFW}
    \rho_{\rm NFW}= \frac{\rho_0 r_s^3}{r (r+r_s)^2}\,,
\end{align}
where $r_s$ is the scale radius and $\rho_0$ sets the overall normalization.
The enclosed mass up to some radius $r$, also known as the halo mass profile, can be computed by
\begin{align}
    M(r)= 4 \pi \int_0^r dr' r'^2 \rho(r')\,.
\end{align}
For an NFW profile the density can be integrated analytically and one finds
\begin{align}
    M_{\rm NFW}(r)=4\pi \rho_0 \, r_s^3 \left[\log(1+\frac{r}{r_s}) -\frac{r}{r_s}\left(1+\frac{r}{r_s} \right)^{-1}  \right]\,.
\end{align} 
As can be seen, $M_{\rm NFW}(r)$ diverges logarithmically for $r\rightarrow \infty$. 
To account for the fact that the halo does not exist in isolation and the DM distribution will be affected by neighboring halos, the spatial extent is conventionally taken to be limited within the virial radius $r_{200}$ which is defined by 
$\bar{\rho}(r_{200})=200 \rho_{\rm crit}$ where $\rho_{\rm crit}$ is the critical density of the universe and $\bar{\rho}({r})$ is the mean density up to this radius.
The viral mass $M_{200}$ of the NFW profile is just $M_{\rm NFW}(r_{200})$. 

The mass and the scale radius of NFW halos are known to be correlated \cite{Navarro:1996gj,Dutton:2014xda}. This removes one of the free parameters from the halo profile such that NFW halos can effectively be characterized by a single parameter. A convenient choice is $M_{200}$. With the help of the concentration parameter $c_{200}$, that has been determined as \cite{Dutton:2014xda}
\begin{align}
    \log_{10} c_{200}=0.905 -0.101 \log_{10} \left(\frac{M_{200} h}{10^{12} M_\odot}\right)
\end{align}
where $h$ is the dimensionless Hubble parameter and $M_\odot$ a solar mass,
one can determine the scale radius via $r_s= r_{200}/c_{200}=\left(\frac{3}{4} \frac{M_{200}}{200 \pi \rho_{\rm crit}}\right)^{1/3} \frac{1}{c_{200}}$. Analogously, the scale density is given by
\begin{align}
    \rho_0= \frac{200 \rho_{\rm crit} c_{200}^3}{3} g_c
\end{align}
with $g_c=\left(\log (1+c_{200}) -\frac{c_{200}}{c_{200}+1} \right)^{-1}$.

The NFW profile is a so-called cuspy profile, i.e.~the density grows $\propto r$ towards the center. Observations of some dSphs are not consistent with this shape of the density distribution and prefer a ``cored" profile instead that is characterized by a core of roughly constant density. We follow \cite{Read:2015sta} and adopt an ansatz for a cored profile that is motivated by the results of N-body simulations that include a modeling of astrophysical feedback. We disregard the possibility of an incomplete cusp-core transition and take \cite{Read:2015sta}
\begin{align}
\label{eq:rho_core}
    \rho_c(r)= \tanh\left(\frac{r}{r_c}\right) \rho_{\rm NFW} + \frac{1- \tanh\left(\frac{r}{r_c} \right)^2}{4\pi r^2 r_c} M_{\rm NFW} (r)\,,
\end{align}
where $r_c$ is the core radius.
For $r\ll r_c$, the density tends to a constant whereas  for $r>r_c$ it just approaches the NFW profile.
This density leads to the simple halo mass profile
\begin{align}
    M_c(r)= \tanh \left( \frac{r}{r_c}\right) M_{\rm NFW}(r)\,,
\end{align}
which transitions quickly to the NFW case for $r>r_c$.

In this work we consider a set of the eight classical Milky Way dSphs: Ursa Minor, Draco, Sculptor, Sextans, Leo I, Leo II, Carina, and Fornax. These were analyzed in \cite{Read:2018fxs} together with a set of irregular dwarf galaxies that is expected to have a more complicated formation history. This reference provides determinations of $M_{200}$ and measurements of the DM density at a fixed radius of 150 pc that are based on fits of stellar kinematics and photometric data with the GRAVSPHERE code \cite{2017MNRAS.471.4541R}. We combine the measurement of $M_{200}$ with the $2\ \sigma$ lower limit on $\rho(150 \,\mbox{pc})$ to derive an upper limit on $r_c$. For convenience we summarize the input from \cite{Read:2018fxs} and the derived quantities in Table \ref{tab:halo}.

\begin{figure}
    \centering
    \includegraphics[width=0.98\textwidth]{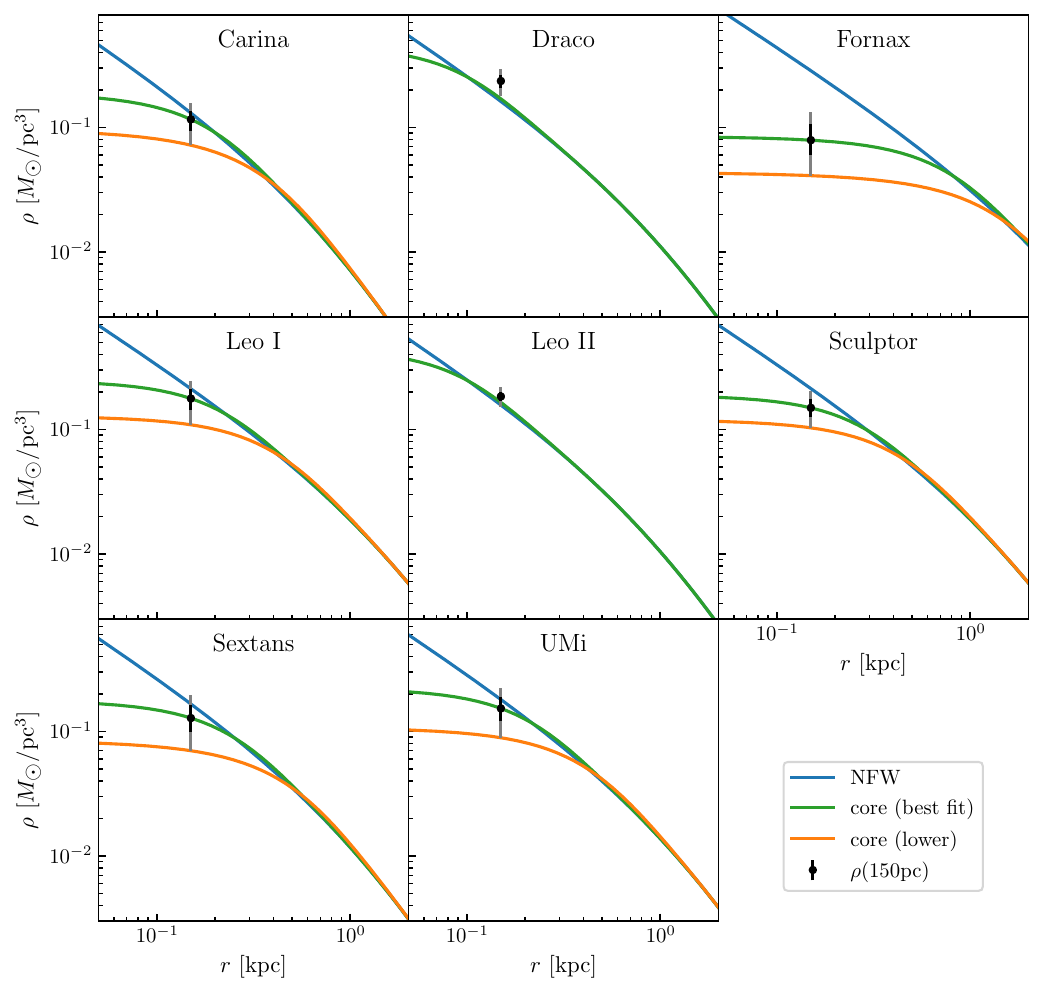}
    \caption{NFW and cored profiles for the dSphs considered in this work. The green curves represent the best-fit cored profiles, which for most cases can reach the central values of   $\rho(150{\rm pc})$, except for Draco and Leo II, for which they only approximately reach the lower bounds of the error bars. The orange line shows the halo with the largest core that is consistent with the measured value of $\rho(150{\rm pc})$.
    } 
    \label{fig:profiles}
\end{figure}

An illustration of the results can be found in Fig. \ref{fig:profiles}. For each dSph we show the NFW profile inferred from $M_{200}$ and the measured value of $\rho_{150}$. Superimposed are the best fit cored profile and the profile with the largest core that is compatible with the measured value of $\rho_{150}$ at $2\ \sigma$. Two of the dSphs (Fornax and Carina) show a preference for a core at more than $2\ \sigma$ while the others are consistent with following an NFW profile on the relevant scales.  Note however that Draco and Leo II do not show any preference for a core since the best-fit density at $150$ pc is higher than the one predicted by the NFW profile. For these we only report the profile with the largest core that does not have a worse agreement with the data than the NFW profile.

\begin{table}[tb]
\centering
\begin{tabular}{|| c | c | c |c || c | c | c | c | c ||} 
\hline
Name  &  $M_{200} [10^{9} M_{\odot}]$  & $M_{*} [10^{6} M_{\odot}]$  & $\rho_{150} [0.1 M_{\odot}/{\rm pc}^3]$   & $\rho_0 [M_\odot/ \mbox{pc}^3]$ & $r_s$ [pc] & $r_{c95}$ [pc] & $\rho_A [M_\odot/ \mbox{pc}^2]$ & $ \Delta E_{\max}[10^{51} \mbox{erg}]$\\[0.5ex]
\hline\hline
Carina & 0.8  & 0.38  & $1.16^{+0.20}_{-0.22}$ & $2.18 \times 10^{-2}$ & $1.14 \times 10^{3}$ & $3.83 \times 10^{2}$ & $4.16 \times 10^{1}$ & $8.57 \times 10^0$ \\[0.5ex]
\hline
Draco$^\ast$ & 1.8 & 0.29 & $2.36{\pm0.29}$ & $1.77 \times 10^{-2}$ & $1.63 \times 10^{3} $ & $ 1.5 \times 10^2$ & $8.0 \times 10^{1} $ & $1.36 \times 10^0$ \\[0.5ex]
\hline
Fornax & 21.9  & 43.0  & $0.79^{+0.29}_{-0.19}$  & $9.39 \times 10^{-3}$ & $ $4.81$ \times 10^{3} $ & $1.56 \times 10^{3} $ & $7.68 \times 10^{1} $ & $1.95 \times 10^{3} $ \\[0.5ex]
\hline
Leo I & 5.6  & 5.5  & $1.77^{+0.33}_{-0.34}$ & $1.32 \times 10^{-2}$ & $2.66 \times 10^{3} $ & $4.1 \times 10^{2} $ & $8.18 \times 10^{1} $ & $3.4 \times 10^{1} $ \\[0.5ex]
\hline
Leo II & 1.6  & 0.74  & $1.84^{+0.17}_{-0.16}$ & $1.82 \times 10^{-2}$ & $1.55 \times 10^{3} $ & $ 1.55 \times 10^2$ & $7.62 \times 10^{1} $ & $1.4 \times 10^0$ \\[0.5ex]
\hline
Sculptor & 5.7  & 2.3  & $1.49^{+0.28}_{-0.23}$ & $1.32 \times 10^{-2}$ & $2.68 \times 10^{3} $ & $4.39 \times 10^{2} $ & $8.02 \times 10^{1} $ & $4.09 \times 10^{1} $ \\[0.5ex]
\hline
Sextans & 2.0  & 0.44  & $1.28^{+0.34}_{-0.29}$ & $1.72 \times 10^{-2}$ & $1.7 \times 10^{3} $ & $5.16 \times 10^{2} $ & $5.13 \times 10^{1} $ & $3.13 \times 10^{1} $ \\[0.5ex]
\hline
Ursa Minor & 2.8  & 0.29  & $1.53^{+0.35}_{-0.32}$ & $1.58 \times 10^{-2}$ & $1.97 \times 10^{3} $ & $4.31 \times 10^{2} $ & $6.28 \times 10^{1} $ & $2.56 \times 10^{1}$\\[0.5ex]
\hline
\end{tabular}
\caption{Halo profile parameters for the dSphs considered here. The parameters $\rho_0$ and $r_s$ are computed from the $M_{200}$ values given by \cite{Read:2018fxs} while the $95\%$ upper limit on $r_c$ has been derived by demanding the $\rho(150 \mbox{pc})$ is consistent with the two sigma lower limit of the same reference. For dSphs marked with $^\ast$ the lower limit of $\rho (150 \mbox{pc})$ is above the value expected for NFW and, therefore, there is no cored halo that fulfills it. We treat this as there being no indication of a core at a radius of $150$ pc and take this as our upper limit. The column density $\rho_A$ has been compute with eq.~\eqref{eq:columnMassDensity} and  using the halo parameters. 
}
\label{tab:halo}
\end{table}

As the total mass of the cored halo is the same while the density at the center is reduced, the cored halo is in an energetically less favorable state than the NFW one. We can estimate the amount of energy required to transform an NFW profile to a cored profile by comparing the total potential energy $W$  of the halos.
It is given by \cite{Binney_and_Tremaine,Penarrubia:2012bb}
\begin{align}
    W= -4 \pi G \int_0^{r_{200}}dr \, r \rho(r) M(r)\,,
\end{align}
where $G$ is Newton's constant.  
The total binding energy of the halo $E$ is related to $W$ by the virial theorem. For the $1/r$ potential of Newtonian gravity, this leads to $E=W/2$. Hence the minimal amount of energy required to transform an NFW to a cored one is  $\Delta E= (W_{c}-W_{\rm NFW})/2 $. 
Therefore, the size of the DM core provides a limit on the amount of energy that can be injected into the DM halo. We use the largest core radius that is consistent with observations to define the energy $\Delta E_{\max}$ that can be absorbed by the halo.  
Typical values are in the range $10^{51}$ to $10^{52}$ erg with Fornax, which has an unusually high virial and stellar mass, an outlier at about  $2\times 10^{54}$ erg, see  Tab. \ref{tab:halo} for a complete list. 
These are large amounts of energy but not exceedingly so when compared to the energy released in astrophysical processes.  A possible source of energy of sufficient order of magnitude that starts to become efficient after the original NFW halo has formed are SN explosion. We will discuss this in the next section.

Before moving on it is worthwhile to consider how robust these results are against variations of the dark matter profile. 
An alternative ansatz for the cored NFW profile suggested by Ref. \cite{Penarrubia:2012bb} is given by
\begin{align}
\label{eq:coredNFW}
    \rho_{c,{\rm alt}}= \frac{\rho_0 r_s^3}{(r+r_c) (r+r_s)^2}\,.
\end{align}
We have repeated our calculation  with this and find that the energy needed to affect a cusp-core transition is a factor of 20-40 larger than what we found using Eq.~\eqref{eq:rho_core}. The reason for this is two-fold: 1) The cored NFW profile of \cite{Penarrubia:2012bb} transitions more slowly towards the core and there are larger differences in the density up to radii of a few $\times r_c$. This means that more material has to be moved further out which requires more energy. 2) In order to  keep the total mass and the virial radius of the halo constant, $\rho_0$ has to be increased for the cored halo compared to the NFW one, see also the discussion in \cite{Penarrubia:2012bb}. This implies that the halo is changed on all scales, even up to $r_{vir}$, and mass is moved to very large radii $r\gg r_s$. The contribution from the very outer parts of the halo is substantial and of a similar order as the one from the changes around $r_c$. This contribution seems artificial to us as we expect the effect of the core formation to be local. The profile of Ref. \cite{Read:2015sta} is motivated by simulations and avoids this ad-hoc rescaling of the halo at large radii. Therefore, we believe that it is a more realistic choice.  Nevertheless, significant astrophysical uncertainty remains, and this should be taken seriously. 

\subsection{Supernova energy}

The energy for reshaping the DM halo can be provided by SN explosions. To get an estimate of the maximal available energy we consider type-II SN explosions only. The total energy released by an explosion is $E_{\rm SN} \approx 3 \times 10^{53}$ erg. In the absence of new physics, $99\%$ of the energy is released in neutrinos. 
According to the ``Raffelt criterion", 
up to an order one fraction of $E_{\rm SN}$  could also be released in the form of light BSM states such as sterile neutrinos, axions, dark photons or the vector bosons of other light new forces.  
To quantify the fraction of SN energy released in this form, we introduce the parameter
\begin{equation}
    \eta \equiv \frac{E_{\rm new}}{E_{\rm SN}}\, , \label{eq:eta}
\end{equation}
where $E_{\rm new}$ denotes the energy released via new particles from a SN.

For the moment we remain agnostic as to the concrete particle species that is produced in the explosion and only want to derive an upper limit on the maximum fraction that can be emitted if an order one fraction of the energy is absorbed by the halo. Therefore, we need an estimate of the total amount of energy that has been released in type-II SN explosions, or, equivalently, the number of SN explosions over the lifetime of the dSphs.

Here we follow \cite{Penarrubia:2012bb} and assume a universal initial mass function (IMF) for the stellar population of the dSphs taken from \cite{Kroupa:2002ky}, known as the Kroupa IMF, which is given by a doubly broken power law with
\begin{align}
    \zeta \propto& \, (m_\ast/m_\odot)^{-0.3}\ \ \mbox{ for } m_\ast \leq 0.08 m_\odot \nonumber\\
       \zeta \propto& \, (m_\ast/m_\odot)^{-1.3}\ \  \mbox{ for } 0.08 m_\odot \leq m_\ast \leq 0.5 m_\odot   \label{eq:IMF-Kroupa}\\
          \zeta \propto& \, (m_\ast/m_\odot)^{-2.3}\ \  \mbox{ for } 0.5 m_\odot \leq m_\ast \nonumber
\end{align}
 where $m_\ast$ is the mass of the star and $m_\odot$ denotes the solar mass.
The continuity of the IMF in Eq.~\eqref{eq:IMF-Kroupa} requires that the coefficients before $(m_\ast/m_\odot)^{-0.3}$, $(m_\ast/m_\odot)^{-1.3}$ and $(m_\ast/m_\odot)^{-2.3}$ should be 1:0.08:0.04, while the overall normalization is unimportant to our calculation. 
Stars in the mass range from $8 m_\odot$ to $50 m_\odot$ have undergone core collapse by now and contribute to the total number of SN explosions over the lifetime of the dSphs. Hence the fraction of stars in this mass range, denoted by $f_{\rm SNII}$, is computed by 
\begin{equation}
    f_{\text{SNII}}=\frac{\int_{8m_{\odot}}^{50m_{\odot}}\zeta(m_{*})dm_{*}}{\int_{0}^{\infty}\zeta(m_{*})dm_{*}}\,.\label{eq:fSN}
\end{equation}
Similar, one can also compute the mean stellar mass  $\langle m_\ast \rangle$ by 
\begin{equation}
    \langle m_\ast \rangle=\frac{\int_{0}^{\infty}\zeta(m_{*}) m_{*} dm_{*}}{\int_{0}^{\infty}\zeta(m_{*})dm_{*}}\,.\label{eq:m-star-mean}
\end{equation}
Using the IMF in Eq.~\eqref{eq:IMF-Kroupa}, we obtain $\langle m_\ast \rangle \approx 0.4 m_\odot$ and $f_{\rm SNII} \approx 3.3 \times 10^{-3}$.  Alternatively, one could also consider the Chabrier IMF~\cite{Chabrier:2003ki},
which for $m_{*}<m_{\odot}$ gives 
$\zeta\propto m_{*}^{-1}\exp\left\{ -\frac{1}{2\sigma_m^2}\left[\log_{10}(m_{*}/m_{c}) \right]^{2}\right\} $
with $m_{c}=\left(0.079_{+0.021}^{-0.016}\right)m_{\odot}$ and $\sigma_m=0.69_{+0.05}^{-0.01}$,
and for $m_{*}>m_{\odot}$ follows the power law $\zeta\propto m_{*}^{-2.35}$.
This leads to $f_{\text{SNII}}\approx2.9\times10^{-3}$
and $\langle m_{*}\rangle\approx0.35m_{\odot}$ and we find that varying the IMF has little impact on our results. We use the Kroupa IMF in our following calculation.

 We estimate the total amount of energy released in SN explosion over the history of the dSphs as 
 \begin{align}
     E_{\rm tot}=E_{\rm SN}\frac{M_\ast}{\langle m_\ast \rangle} f_{\rm SNII}\approx 2.5  \frac{M_\ast}{m_\odot} \times 10^{51}\mbox{erg}\,,
 \end{align}
 where $M_\ast$ is the total stellar mass. We use the values of $M_\ast$ reported in \cite{Read:2018fxs}. 
 By comparing $E_{\rm tot}$ with the maximal energy that can be transferred to the halo $\Delta E_{\max}$ we get an upper limit on the fraction of energy that a type-II SN explosion can release in exotic particles.

\begin{figure}
     \centering
     \includegraphics[width=0.8\textwidth]{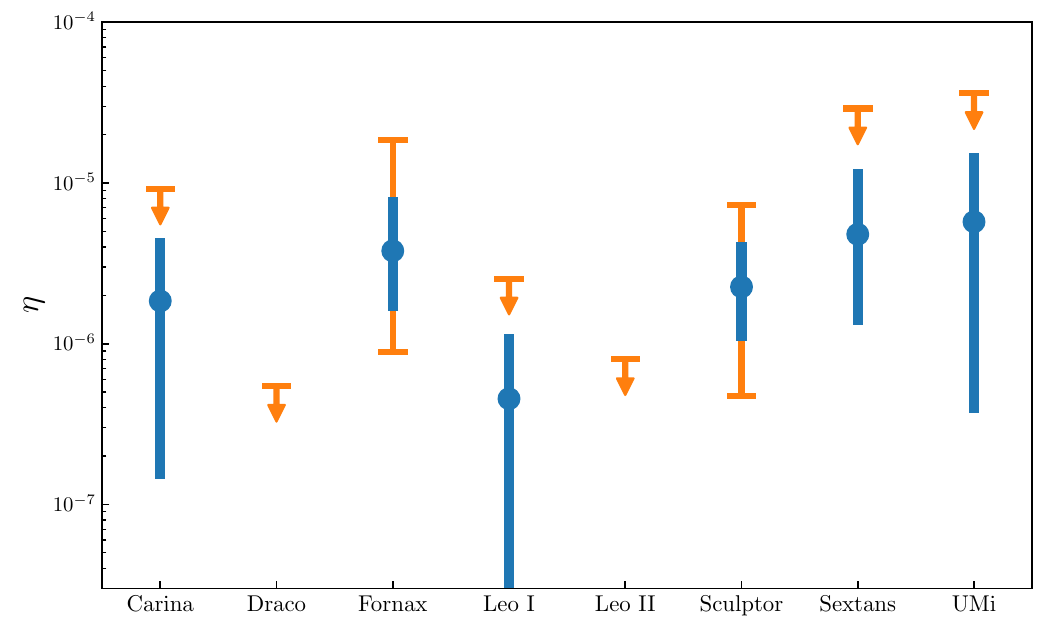}
     \caption{Upper limit on the fraction of energy that can be released in exotic particles if it is absorbed by the DM halo afterwards for a set of eight classical dSphs. The limits are derived from the data and the halo profile suggested in \cite{Read:2018fxs}  with an upper limit on $r_c$ determine from the upper limit on $\rho_{150}$. 
     For Draco and Leo II which do not favor the cored profile, we only set the upper bounds on $\eta$ by requiring $r_c\leq 0.095$ kpc and $0.158$ kpc, respectively.  
     Blue bars indicate 1$\sigma$ intervals favored by observations; orange bars or arrows indicate 2$\sigma$ intervals or upper bounds. 
     } 
     \label{fig:exotics_fraction}
 \end{figure}

 A summary of the results can be found in Fig.~\ref{fig:exotics_fraction}.
As can be seen, the preferred range for the energy to be injected into the DM halo is about $10^{-6}$ to $10^{-5}$ of the total energy released in the explosion. In addition, none of the considered dSphs is consistent with a core size significantly above the one that corresponds to an energy injections larger than a few$\times 10^{-5}$ of the energy released by SN explosions. 
Thus only a small amount of energy can be absorbed by the halo directly. These results can be interpreted in a number of ways. First, it is clear that a large fractional energy release is at odds with the observed properties of the halo if the efficiency with which the energy is absorbed is $\mathcal{O}(1)$. 
This allows to place an upper limit on the efficiency of exotic particle production which can be interpreted as an upper limit on the interaction strength in specific particle physics models. 
We will mainly follow this line of thought in the following and investigate the conditions for this situation from a particle physics perspective in the next two sections. Second, it remains possible that a significant fraction of the energy is released into the dark sector if the efficiency of absorbing is small. 
Nevertheless, even in this case, interesting effects can appear if the efficiency is in the range from $10^{-4}$ to $10^{-1}$, depending on the fraction of energy that goes into new physics states. This regime covers the transition from an opaque to an almost transparent halo. Modeling radiation transport over such a large range of opacities is tricky and we leave more detailed considerations of this possibility for future work. Finally, as mentioned above, some of the dSphs show a preference for a core that is not expected to form based on DM-only simulations. 
This shortcoming can be alleviated if the energy required for the core formation is provided by dark radiation from SN explosions. Note, however, that
simulations that try to take baryonic feedback into account, show formation of cores for certain parameters of the gas model, see e.g. \cite{Read:2015sta}. The mechanism at work there has some similarities to the one considered here, in that the effect on the halo can be explained by energy injection from SN explosions. In this scenario, the energy is transmitted to the DM halo via gravitational interaction between regular matter expelled by the explosion and the DM particles that make up the halo. 
Unfortunately, the interaction strength with the SM that leads to a fractional energy release in the range $10^{-6}$ to $10^{-5}$ is very small which makes testing it in direct experiments very challenging. One might speculate that more detailed observations of the neutrino emission from a future galactic SN could provide new insights here.
Finally, considering the alternative  halo profile in Eq.~\eqref{eq:rho_core} relaxes the limit on $\eta$ since more energy is required to create a sizable core in this case. The upper limits for Fornax, which clearly prefers a core, shifts to  $3. \times 10^{-4}$ while Carina, Leo I, Sculptor, Sextans and Ursa Minor lead to upper limits with similar values. For Draco and Leo II, $r_c$ needs to be chosen significantly smaller than $150 \mbox{pc}$ since the $2\sigma$ lower limit on $\rho(150 \mbox{pc})$ is already larger than the NFW value.  This leads to values of $\eta$ of a few $\times 10^{-5}$.  However, the choice of $r_c$ is somewhat arbitrary here. 
 
\section{Particle physics estimates}
\label{sec:particle}

As we have seen in the previous section, if SN explosions in a galaxy  deposit
a small fraction of their total energy into the DM halo, it may significantly
affect the structure of the DM halo. In this section, we investigate 
this possibility from the perspective of particle physics. 
We keep the discussion relatively generic here and provide the basic ingredients that are needed to identify the requirements on the particle physics properties of
DM and auxiliary particles. 
We consider a template model in which a massive vector boson $Z'$ with free couplings to light fermions (referred to as the dark radiation in the following) drains a small fraction of energy from the explosion and deposits it into the DM halo.
We estimate the emissivity of such particles, and investigate 
the effectiveness of their energy being absorbed by the DM halo.

Concretely, the interaction Lagrangian of our template model reads
\begin{equation}
{\cal L}_{\rm int}\supset g_{\chi}\overline{\chi}Z'_{\mu}\gamma^{\mu}\chi+\sum_{\psi=e,n,p,\nu,\cdots}g_{\psi}\overline{\psi}Z'_{\mu}\gamma^{\mu}\psi\thinspace,\label{eq:L}
\end{equation}
where $\chi$ is the DM particle, $\psi$ denotes fermions that are
present in the SN medium, i.e.~neutrinos, electrons, muons proton, and neutrons. 
Before going into any more details, we want to comment that the rates for the production of a massive vector share many properties with a scalar and the difference between these rates is typically a factor of two, see e.g.~\cite{Raffelt:1996wa,Bottaro:2023gep}. 
Therefore, our analyses below can be applied to the scalar case as well up to such variations\footnote{In contrast, the rates for light pseudoscalars, e.g.~the axion, are known to deviate significantly in certain kinematic regimes and an application to this case would require a more detailed, dedicated analysis.}. We want to emphasize here that our analysis is only supposed to identify the right order of magnitude and does not aim to compete with a complete dedicated analysis of particle production in a concrete model.

\subsection{Production rates\label{subsec:Dominant-production}}

Let us first compute the production rate of $Z'$ in the presence
of the generic couplings in Eq.~\eqref{eq:L}.  In the hot and
dense plasma of a SN core, a number of processes can contribute significantly to the production of light new bosons. We restrict ourselves to the subset of processes that dominate the
production in at least one of the models we consider in Sec.~\ref{sec:models}. 
For simplicity we consider only one process per constituent of the core:
\begin{itemize}
\item Nucleon bremsstrahlung (NBr): $N+N\to N+N+Z'$; 
\item Semi-Compton scattering (SC): $\gamma+e^{-}\to Z'+e^{-}$; 
\item Semi-Compton muon scattering (SC-$\mu$): $\gamma+\mu^{\pm}\to Z'+\mu^{\pm}$; 
\item Neutrino coalescence ($\nu$Co): $\nu+\overline{\nu}\to Z'$. 
\end{itemize}
Among these processes, NBr is important to $Z'$ with sizable
hadronic couplings since the nucleon scattering cross section is very large. For a $Z'$ with
couplings to charged leptons, SC is the most relevant process. 
In addition, such a $Z'$  can also be produced via electron bremsstrahlung ($e^{-}+N\to e^{-}+N+Z'$), which in the SN core is subdominant compared to SC. 
In the Sun and red giants, however, electron bremsstrahlung can be a dominant channel---see e.g.~\cite{Li:2023vpv}.
It is noteworthy that SC on muons, which
may be present in the SN with non-negligible abundance~\cite{Bollig:2020xdr},
  could  be the dominant production channel for muonphilic $Z'$~\cite{Croon:2020lrf}.
The last channel, $\nu$Co, is important to
neutrinophilic radiation such as the Majoron~\cite{Chikashige:1980ui}
or dark $Z'$ arising from the right-handed neutrino sector~\cite{Chauhan:2020mgv,Chauhan:2022iuh}. 
Neutrinophilic $Z'$ may also be produced via neutrino bremsstrahlung: $\nu+N\to\nu+N+Z'$. We have estimated the production rate of this process and find that its contribution is negligible.

{\bf NBr:} Two nucleons interact with each other mainly via strong
interactions. This greatly enhances the cross section in comparison
to photon-mediated processes such as electron bremsstrahlung and makes NBr an important production channel if the nucleon couplings are not suppressed. 
It is important to note that if the two nucleons are identical (such as in a proton-proton or neutron-neutron collision), the dipole emission rate of $Z'$ vanishes, leaving the quadrupole emission as the dominant contribution---see e.g.~Refs.~\cite{Rrapaj:2015wgs,Dev:2020eam} for discussions. The dipole emission is also approximately cancelled out in neutron-proton collision if $Z'$ is equally coupled to the two different nucleons.
For $Z'$ with different couplings to the neutron and the proton, both dipole and  quadrupole contributions are present, with the former larger than the latter typically by a factor of 5 to 7~\cite{Rrapaj:2015wgs}.
Depending on whether the production is dominated by dipole or quadrupole emission, we further categorize the processes as NBr-2 and NBr-4, respectively. 

For NBr-2,  the production rate is computed by~\cite{Chang:2016ntp}
\begin{equation}
\Gamma_{\text{NBr}}=e^{-\omega/T}\frac{32\alpha_{N}'}{3\pi\omega^{3}}\left(\frac{\pi T}{m_{N}}\right){}^{3/2}n_{n}n_{p}\langle\sigma_{np}\rangle\xi_{\rm TL}\thinspace,\label{eq:-4}
\end{equation}
where $\omega$ is the energy of the $Z'$, $T$ is the temperature of the core, $m_N$ is the nucleon mass,  $\xi_{\rm TL}=1$ or $m_{Z'}^{2}/\omega^{2}$ for transverse and
longitudinal polarizations, and $\langle\sigma_{np}\rangle$ is the
thermally averaged proton-on-neutron scattering cross section. Throughout this work, we denote the density of particle species $i$ by $n_i$ and define $\alpha_i'\equiv g_i^2/4\pi$, which is the equivalent of the fine-structure constant for the $Z'$ coupling to that particle species.

One might be tempted to compute
NBr by considering the pion as a mediator between two nucleons which leads to the one-pion-exchange potential.  However,
it has been shown that, for nucleon cross sections at the energies
considered here,
this is
not a good approximation and an improved cross section should be used~\cite{Rrapaj:2015wgs}. We extract $\langle\sigma_{np}\rangle$ from Fig.~6
of said reference. 
For NBr-4, we use a formula similar to Eq.~\eqref{eq:-4} except that  $\langle\sigma_{np}\rangle$ is replaced by the quadrupole cross section which is also available from Ref.~\cite{Rrapaj:2015wgs}.

{\bf SC:} The production rate for the SC process can be computed in terms of the Klein-Nishina cross section via (see, e.g., \cite{Redondo:2013wwa}) 
\begin{equation}
\Gamma_{\text{SC}}=e^{-\omega/T}\sigma_{T}\frac{\alpha_{e}'}{\alpha}F_{\text{rel}}F_{\text{deg}}n_{e}\xi_{\rm TL}\thinspace,\label{eq:-6}
\end{equation}
where $\sigma_{T}=\frac{8\pi\alpha^{2}}{3m_{e}^{2}}$ is the Thomson
cross section, $F_{\text{rel}}$ is the relativistic correction factor, and $F_{\text{deg}}$ is another factor accounting
for the  degeneracy of the electron gas \cite{Raffelt:1996wa}. 
The Klein-Nishina  relativistic correction
factor reads\footnote{This is for example given in Eq.~(5-116) in \cite{Itzykson:1980rh}. This factor can be obtained by analytically calculating the total cross section using the Klein-Nishina formula. }
\begin{equation}
F_{\text{rel}}(x)=\frac{3}{4}\left(\frac{(x+1)}{x^{3}}\left[\frac{2x(x+1)}{2x+1}-\log(2x+1)\right]+\frac{\log(2x+1)}{2x}-\frac{3x+1}{(2x+1)^{2}}\right),\label{eq:-26}
\end{equation}
where $x\equiv\omega/m_{e}$. The degeneracy factor $F_{\text{deg}}$
can be estimated by averaging the Pauli blocking factor~\cite{Raffelt:1996wa} 
\begin{equation}
F_{\text{deg}}=\frac{2}{n_{e}}\int\frac{{\rm d}^{3}\mathbf{p}}{(2\pi)^{3}}f_{e}\left(1-f_{e}\right),\label{eq:-27}
\end{equation}
where $f_{e}$ is the phase space distribution of the electron. In
the degenerate limit, $F_{\text{deg}}$ is approximately given by
$F_{\text{deg}}\approx3E_{F}T/p_{F}^{2}$ where $p_{F}$ and $E_{F}$
are the momentum and energy of the electron at the Fermi surface.
In the SN core where  $E_{F}\gg m_{e}$, taking typical core values of the temperature $T\sim 30 {\rm MeV}$ and the density $\rho\sim 10^{15}{\rm g}/{\rm cm}^3$, this value gives $F_{\text{deg}}\approx3T/p_{F}\sim 0.3$. 
In the non-relativistic
non-degenerate limit, Eqs.~\eqref{eq:-26} and \eqref{eq:-27} reduce
to $F_{\text{rel}}\approx F_{\text{deg}}\approx1$.

{\bf \boldmath SC-$\mu$:} The results for SC on electrons can be straightforwardly generalized to calculate the
SC-$\mu$ process. This only requires replacing $n_{e}\to n_{\mu}$ , $m_{e}\to m_{\mu}$,
, $\alpha_{e}'\to\alpha_{\mu}'$, and recomputing $F_{\text{deg}}$ for the muon case. It typically varies from 1 to 0.85 (see Fig.~7 of \cite{Bollig:2020xdr}) hence is neglected in our calculation of SC-$\mu$.

{\bf \boldmath $\nu$Co:} We assume that the neutrino ($\nu$) and antineutrino
($\overline{\nu}$) phase space distributions are given by $f_{\nu}\approx\exp\left[-(E_{\nu}-\mu_{\nu})/T\right]$
and $f_{\overline{\nu}}\approx\exp\left[-(E_{\overline{\nu}}-\mu_{\overline{\nu}})/T\right]$,
with opposite chemical potentials: $\mu_{\overline{\nu}}=-\mu_{\nu}$.
Correspondingly, the ratio of their number densities is $n_{\nu}/n_{\overline{\nu}}\approx e^{2\mu_{\nu}/T}$.
Although $n_{\nu}$ is much higher than $n_{\overline{\nu}}$ during
the \emph{neutronization}, the production rate of $Z'$ relies on
$f_{\nu}f_{\overline{\nu}}$, in which the two chemical potentials
cancel out.\footnote{We note here that for the Majoron or other particles with 
lepton number violation, this cancellation is absent since
the production rate relies on $f_{\nu}f_{\nu}$ rather than  $f_{\nu}f_{\overline{\nu}}$.
In this case, the production would be much more efficient.} As a consequence, the production rate of $Z'$ via $\nu$Co only
depends on the local temperature (see, e.g., Eq.~(A.24) in Ref.~\cite{Li:2022bpp}): 
\begin{equation}
\Gamma_{\text{\ensuremath{\nu}Co}}\approx\frac{\alpha_{\nu}'m_{Z'}^{2}}{4\pi\omega}e^{-\frac{\omega}{T}}\thinspace.\label{eq:-30}
\end{equation}

With the dominant production rates presented above, we compute the
luminosity of the SN core by 
\begin{equation}
L_{Z'}\equiv\int_{0}^{R_{c}}dr4\pi r^{2}\int_{0}^{\infty}dk\frac{k^{2}}{2\pi^{2}}\omega\Gamma_{{\rm prod}}\left\langle \exp\left[-\tau_{{\rm SN}}(r,k)\right]\right\rangle \thinspace,\label{eq:-12}
\end{equation}
where $k$ is the momentum of the $Z'$, $R_{c}$ is the core radius,
$r$ is the distance to the center, and $\Gamma_{{\rm prod}}$ represents
the sum of the relevant production rates. In Eq.~(\ref{eq:-12}),
we introduce the SN optical depth $\tau_{{\rm SN}}$ to account for
potential trapping of $Z'$ in the SN. To take the direction of the
outgoing $Z'$ into account, we calculate the directional average
of $e^{-\tau_{{\rm SN}}}$  following~\cite{Caputo:2021rux}
\begin{align}
\left\langle \exp\left[-\tau_{{\rm SN}}(r,k)\right]\right\rangle  & =\frac{1}{2}\int_{-1}^{+1}d\cos\theta\thinspace\exp\left\{ -\int dl\left[\Gamma_{\text{decay}}\left(r',k\right)+\Gamma_{\text{scat}}\left(r',k\right)\right]\right\} ,\label{eq:avg-exp}
\end{align}
with
\begin{equation}
r'=\sqrt{r^{2}+l^{2}+2r\thinspace l\thinspace\cos\theta}\thinspace,\label{eq:rp-r-l}
\end{equation}
where $\theta$ is the angle of the outgoing $Z'$ with respect to
the radial direction, $l$ is the flight distance, and $\Gamma_{\text{decay}}$
and $\Gamma_{\text{scat}}$ denote the decay and scattering rates
of $Z'$ in the SN medium, respectively. 
For $Z'$ decaying to neutrinos, the effective contribution should
be limited within the neutrinosphere with the radius $R_{\nu}\approx30$
km---see Ref.~\cite{Chang:2016ntp} for further details. For $Z'$
decaying to other SM particles, a far radius $R_{{\rm far}}\approx100$
km is used instead of $R_{\nu}$~\cite{Chang:2016ntp}. 

For heavy $Z'$ with sizable $g_{\chi}$, it may dominantly decay to $\chi$. In this case, 
one might be concerned
about the opacity of SN to $\chi$'s produced in the core. For example,
$\chi$ could scatter with nucleons in the core via the $t$-channel
process $\chi N\to\chi N$ mediated by a $Z'$, and could get
trapped too.  This process is suppressed by an additional coupling, but one would expect this suppression to be mild  since the  couplings needed to realize the energy transfer mechanism will turn out to be sizable\,---\,see Sec.~\ref{subsec:Lifetime-opacity} for more details. Therefore, in the regime where the coupling connecting the dark and the visible sector is not very small anymore,  DM trapping is a plausible possibility. This effect further complicates the computation of the energy loss and would require a dedicated analysis. In this work we are mainly interested in possible improvements on the cooling bound at small couplings where these effects are not expected to be relevant. We therefore leave a detailed study of dark matter trapping for future work and caution the reader that our arguments should be taken with a grain of salt in the trapping regime.

The calculation of Eq.~\eqref{eq:-12} requires a specific SN profile to be used in the integral.
We adopt a simulated profile from Ref.~\cite{Bollig:2020xdr}, which
 allows us to  take the muon number density into account consistently
with  other factors such as the temperature and the densities of other medium particles.
More specifically, we use the SFHo-18.6 model 
and compute the proton and neutron number densities by $n_{p}=Y_{p}\rho/m_{N}$
and $n_{n}=(1-Y_{p})\rho/m_{N}$ with $Y_{p}\approx0.3$ and $\rho$
the matter density. The electron number density is determined by the
electric neutrality of the medium: $n_{e}=n_{p}$. The neutrino number
density is determined by assuming that the lepton number is approximately
conserved within the neutrino sphere. This leads to $n_{\nu}-n_{\overline{\nu}}\approx n_{e0}-n_{e}\approx\left(1/2-Y_{p}\right)\rho/m_{N}$,
where $n_{e0}$ denotes the value of $n_{e}$ before the collapse.
The muon number density can be obtained from Fig.~3 of~\cite{Bollig:2020xdr}. 
Note that strictly speaking, the core profile is not static and
varies with time, causing temporal variation of $L_{Z'}$. In principle,
one should integrate a dynamic profile over time to get the total
energy emitted in the new physics channel (see, e.g., Ref.~\cite{Caputo:2022mah})
and then compare it with $E_{\text{SN}}$ to obtain $\eta$. Here
we take a simplified treatment assuming  $E_{{\rm new}}\sim L_{Z'}\Delta t$,
with $L_{Z'}$ obtained using the above static profile and $\Delta t\sim10$
s. We expect that $E_{{\rm new}}$ computed in this way may deviate
 from the true value by a factor of a few, mainly due to the relatively
mild variation of the core temperature---see Fig.~11.2 of \cite{Raffelt:1996wa}.

\begin{figure}
\centering

\includegraphics[width=0.7\textwidth]{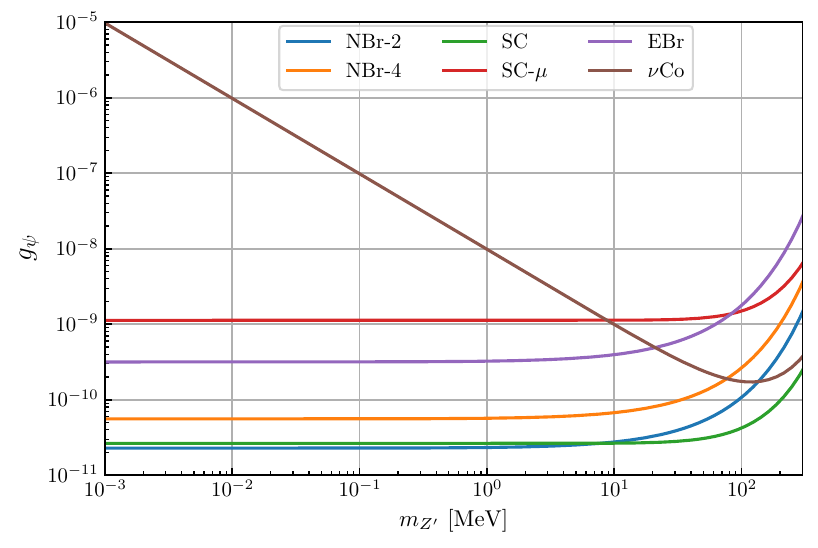}

\caption{
The required in medium coupling strength of $Z'$ to generate $L_{Z'}=3\times10^{52}\ \text{erg}/\text{sec}$, assuming the SN optical depth $\tau_{\rm SN}=0$. 
  \label{fig:g_bound}}
\end{figure}


The result is illustrated in Fig.~\ref{fig:g_bound} in terms of the
required coupling strength $g_{\psi}$ to generate $L_{Z'}=L_{\nu}$,
where $L_{\nu}\equiv 3 \times10^{52}\ \text{erg}/\text{sec}$ is 
the SN neutrino luminosity in average, assuming that $E_{\rm SN}\approx 3\times 10^{53}\ {\rm erg}$ is released in about 10 seconds.
As can be seen
from Fig.~\ref{fig:g_bound},  NBr and SC are typically the most
efficient production channels for $Z'$ universally coupled to all
fermions. Other channels may be important if the model features drastically different couplings to the light fermions. The impact of such a variation for  representative set of benchmark models  will be discussed in Sec.~\ref{sec:models}.

\subsection{Lifetime, column density, and opacity\label{subsec:Lifetime-opacity}}

\begin{figure}
\centering

\includegraphics[width=0.95\textwidth]{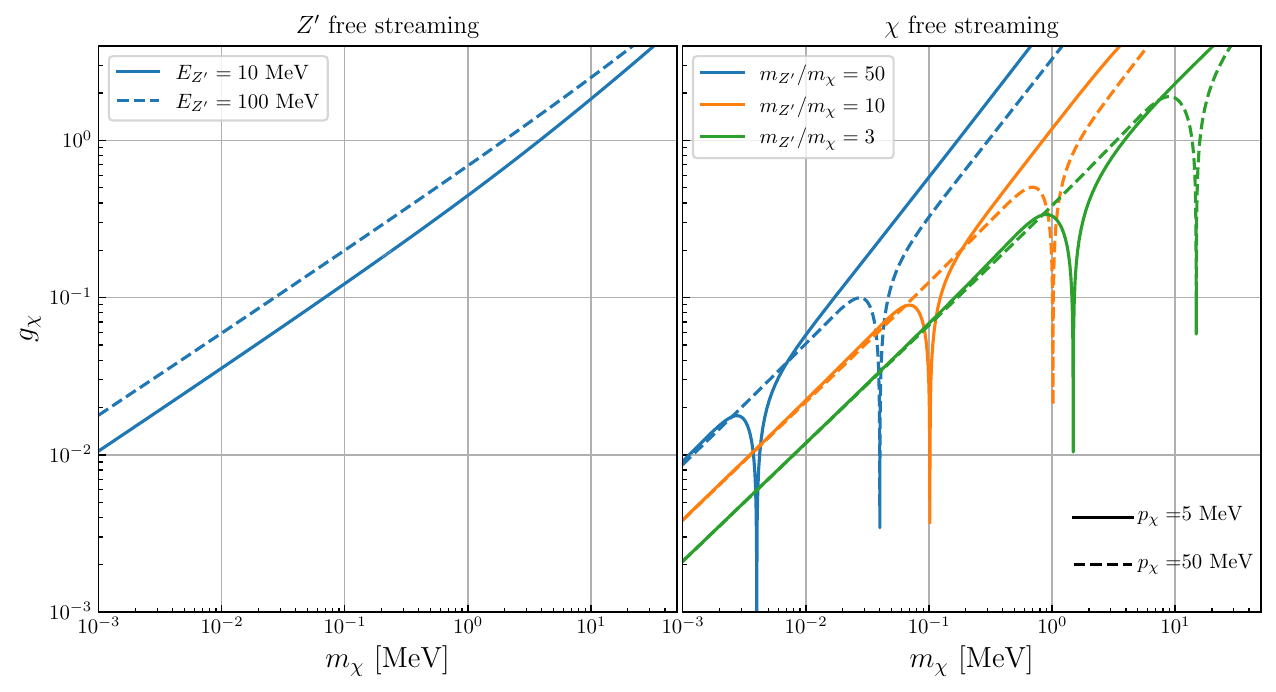}

\caption{Required coupling strengths for SN emitted particles to effectively
deposit their energy into the DM halo. The left and right panels concern the optical depth of the halo for $Z'$ and $\chi$, respectively.
\label{fig:g-chi}
}
\end{figure}

We are interested in the situation where the energy released in dark radiation by SN explosions is largely deposited into the DM halo.
While we do not attempt to model the details of the energy transfer, we want to identify the conditions that are required for a successful energy transfer. Here, the first question is
 whether the $Z'$ particles are stable or not on the relevant
astrophysical distances.
More specifically, if a $Z'$ particle is
not absolutely stable, the mean distance that it can travel before decaying is
\begin{equation}
l_{{\rm decay}}=\frac{1}{\Gamma}\gamma \beta\thinspace,\label{eq:-23}
\end{equation}
where  $\Gamma$ is the width of the particle, 
$\beta$ its velocity and $\gamma$ the Lorentz factor. 

We need to differentiate between two possibilities now. On the one hand,
if $l_{{\rm decay}}$ is much longer than the size of the DM halo,
we consider it as practically stable. In this case, the major concern
is whether its cross section with non-relativistic DM particles can be  large enough to
make the DM halo opaque to $Z'$.
On the other hand,
if $l_{{\rm decay}}$ is much shorter than the size of the DM halo,
it will loss its energy to the daughter particles in the decay and not to the halo directly. In this case, the major concern
is whether it dominantly decays to dark-sector particles (e.g., to
a DM pair directly) and whether the DM halo is opaque to the energetic
decay products. 

The first step to address these questions is to estimate
the lifetime of $Z'$. Although the quantitative calculations of the
lifetime is model dependent, we can still obtain some generic results
that  indicate at least the correct order of magnitude.
Anticipating the results for the opacity of the DM halo (to be computed later and presented in Fig.~\ref{fig:g-chi}), we expect that the coupling $g_\chi$ has to be sizable if the energy transfer to the halo is efficient. In contrast, the couplings to the SM that lead to a luminosity lower than the SM neutrino one are usually very small. Therefore, the branching ratio and the lifetime of the $Z'$ depends crucially on the ratio $m_{Z'}/m_\chi$. For $m_{Z'}/m_\chi> 2$, decays to two DM particles are kinematically allowed. Due to the large coupling and the absence of further suppressing factors $l_{\rm decay}$ will be microscopic and the branching ratio to DM  $100\%$ for all practical purposes. For $m_{Z'}/m_\chi \leq 2$, this decay is not possible and the $Z'$ has to decay to SM states. 
In this case the situation is less clear and macroscopic decay length are possible, which will be estimated in the following.

First, in the presence of $g_{\nu}$ the contribution of a single neutrino species to the
decay width of $Z'$,   which can be straightforwardly computed according to Eq.~\eqref{eq:L}, reads:
\begin{equation}
\Gamma_{Z'\to2\nu}\approx\frac{g_{\nu}^{2}m_{Z'}}{24\pi}\thinspace.\label{eq:-38}
\end{equation}

To compute $l_{{\rm decay}}$, we also need $\beta\gamma=p_{Z'}/m_{Z'}$ where $p_{Z'}$ is the momentum of $Z'$. The distribution of $p_{Z'}$ depends on specific production processes. For instance, the NBr process typically tend to produce relatively soft $Z'$ due to the $\omega^{3}$ factor in the denominator of Eq.~\eqref{eq:-4}, while the SC process tend to produce $Z'$ with a harder spectrum. Given that the core temperature is a few tens of MeV, we expect typical values of $p_{Z'}$ in the range 10 to 100 MeV.

Taking Eq.~\eqref{eq:-23} with Eq.~\eqref{eq:-38}, we obtain
\begin{equation}
l_{{\rm decay}}\approx\ 1\ \text{kpc}\cdot\left(\frac{0.4\times10^{-10}}{g_{\nu}}\right)^{2}\cdot\left(\frac{0.1\ \text{keV}}{m_{Z'}}\right)^{2}
\cdot\left(\frac{p_{Z'}}{30\ \text{MeV}}\right)
\thinspace.\label{eq:-39}
\end{equation}

Second, a similar estimate can also be applied to $Z'\to2e$, provided that
$m_{Z'}$ is significantly above $2m_{e}\approx1$ MeV. In this case,
we rescale the benchmark mass $0.1$ keV in Eq.~\eqref{eq:-39} by
at least a factor of $10^{5}$. Correspondingly, the benchmark value
of the coupling would be decreased by at least a factor of $10^{5}$.
Therefore, we conclude that for $m_{Z'}$ significantly above $1$
MeV, $l_{{\rm decay}}\lesssim1\ \text{kpc}$ requires the coupling
$g_{e}$ to be below $5\times10^{-16}$. This is much lower than
any of the typical values presented in Fig.~\ref{fig:g_bound}, implying
that in general $Z'$ above $2m_{e}$ cannot be stable on
astrophysical scales relevant to our work, unless its coupling $g_{e}$
is highly suppressed compared to other effective couplings (see the
example of dark Higgs to be discussed later). 

Finally, if couplings to $\nu$ are absent and $m_{Z'}\leq 2 m_e$, the lowest multiplicity final states are $\gamma \gamma$ for a scalar or $3\gamma$ for a vector. As there is no tree-level coupling to photons in either case these decays are loop induced and the width is model-dependent. We will therefore postpone a more detailed discussion of these decays until Sec.~\ref{sec:models}. 

The next step is to estimate the opacity of the halo to the dark radiation
produced from SN explosions directly or its decay product. The general formula
for computing the optical depth is given by 
\begin{align}
\tau=\langle\sigma v\rangle\,\frac{\rho_{A}}{m_{\chi}}\thinspace,
\end{align}
where $\langle\sigma v\rangle$ is the appropriate average of the
scattering cross section times velocity, which reduces to $\sigma$
for relativistic particles, and $\rho_{A}$ denotes the column mass
density defined by 
\begin{align}
\rho_{A}=\int\rho dl\thinspace,\label{eq:columnMassDensity}
\end{align}
where $l$ is the way to the edge of the halo. For $r_{c}<r_{s}$
and $r_{s}\ll r_{vir}$ the result for the cored profile is to better
than $5\%$ precision approximated by 
\begin{align}
\rho_{A}\approx\rho_{0}r_{s}\left(\log\frac{r_{s}}{r_{c}}+\frac{1}{2}\right).
\end{align}
We report the column density from a full numerical integration in
Tab.~\ref{tab:halo}. If the particles that are emitted by the SN
explosion move with relativistic velocities, we can get a simple estimate
of the cross section that is required to transfer an order one fraction
of the energy to the halo by requiring $\tau >1$ or, equivalently, $\sigma\gtrsim m_{\chi}/\rho_{A}$. Taking
the values of $\rho_{A}$ in Tab.~\ref{tab:halo}, we find that this
corresponds to 
\begin{equation}
    \sigma\gtrsim (1.0 \-- 2.1) \times10^{-25}\mbox{cm}^{2}\cdot\left(\frac{m_{\chi}}{\text{MeV}}\right).\label{eq:-41}
\end{equation}
In the following analysis, we take $\sigma\gtrsim1.0\times10^{-25}\text{cm}^{2}$
as the requirement of the opacity. If the larger value is used, the corresponding value of $g_{\chi}$
required by the opacity would change by a factor of $2.1^{1/4}\approx1.2$
according to Eqs.~\eqref{eq:-40}-\eqref{eq:-40-chichi} below.

For stable $Z'$, we are concerned about the DM halo opacity to $Z'$s. In the
 simplest scenario they would loose their kinetic energy to the halo
via $Z'\chi\to Z'\chi$ scattering. Taking $m_{Z'}\ll m_{\chi}$ for simplicity, the total cross section of the process 
is given by the  Klein-Nishina formula, according to which we obtain
\begin{equation}
\sigma_{Z'\chi\to Z'\chi}\approx\frac{8\pi\alpha_{\chi}^{2}}{3m_{\chi}^{2}}F_{\text{rel}}\left(\frac{E_{Z'}}{m_{\chi}}\right),\label{eq:-40}
\end{equation}
where $\alpha_{\chi}=g_{\chi}^{2}/(4\pi)$ and $F_{\text{rel}}$ has
been given by Eq.~\eqref{eq:-26}. 
By requiring that $\sigma_{Z'\chi\to Z'\chi}$ is above the lower
bounds in Eq.~\eqref{eq:-41}, we obtain the corresponding lower bounds
on $g_{\chi}$, which are presented in Fig.~\ref{fig:g-chi}. As can be seen, we require $m_\chi \lesssim $ a few MeV and $g_{\chi}$ in the range $0.01$ to $1$ in order to make the halo opaque to the dark radiation.

For unstable $Z'$s, we concentrate on the case that $Z'\to2\chi$
dominates the decay. This can be achieved easily if the decay is kinetically allowed and the coupling to the DM is  larger than the values of $g_{\psi}$ presented in
Fig.~\ref{fig:g_bound}. So the
major concern becomes whether the energetic (typically relativistic,
in contrast to the $\chi$ particles in the halo) $\chi$ particles
produced from $Z'\to2\chi$ could deposit the bulk of
their kinetic energy in the halo. 
The simplest process, that can proceed  without introducing other interactions, is elastic
scattering between DM particles mediated by a $Z'$. It receives contributions from $\chi \bar{\chi}$ and  $\chi \chi$ scattering.  We compute the cross section for both cases including all masses. 
In the limit $m_\chi^2\ll s$, our result reduces to
\begin{equation}
\sigma_{\chi\overline{\chi}\to\chi\overline{\chi}}=\frac{g_{\chi}^4}{4\pi}\frac{1}{(s-m_{Z'}^{2})^2}\left( \frac{s}{3} + \frac{s^2}{m_{Z'}^{2}} -\frac{2 m_{Z'}^{4}}{m_{Z'}^{2}+s} + \frac{2(m_{Z'}^{2}-s)(m_{Z'}^{2}+s)}{s} \log\left[\frac{m_{Z'}^{2}+s}{m_{Z'}^{2}}\right]\right),
\end{equation}
which features a resonance when the Mandelstam variable $s$ is approaching
$m_{Z'}^{2}$ as expected. The resonance can be regulated by a Breit-Wigner ansatz when necessary. For 
$\chi\chi\to\chi\chi$ scattering,  we find
\begin{equation}
\sigma_{\chi \chi\to\chi \chi}=\frac{g_\chi^4}{4\pi} \frac{1}{s}\left(1+\frac{2 s}{m_{Z'}^{2}}+ \frac{m_{Z'}^{2}}{m_{Z'}^{2}+s}-m_{Z'}^{2}\left(\frac{1}{s}+\frac{2}{2m_{Z'}^{2} +s} \right)\log \left[\frac{m_{Z'}^{2}+s}{m_{Z'}^{2}} \right] \right)\,.
\label{eq:-40-chichi}
\end{equation}
When estimating the optical depth, we take the full expressions including the $m_\chi$ mass dependence and average the two cross sections since we assume that the DM halo consist of equal numbers of $\chi$ and $\bar{\chi}$.\footnote{An asymmetric DM scenario where only one of the two is present in the halo only leads to minor changes in the averaging since the SN explosion produces equal amounts of $\chi$ and $\bar{\chi}$ in the models considered here.}
By requiring that the combined cross section is above the lower bounds
in Eq.~\eqref{eq:-41}, we obtain the corresponding lower bounds on
$g_{\chi}$, as shown in Fig.~\ref{fig:g-chi}.
In the shown examples, 
we fix $p_{\chi}$ and $m_{Z'}/m_{\chi}$ at a few representative values indicated in the figure. The dips
on these curves are caused by the $s$-channel resonance. 
The results in Fig.~\ref{fig:g-chi} suggest that the required magnitude
of $g_{\chi}$ for $Z'$ or its decay product $\chi$ to fully deposit
the energy into the halo typically varies from $10^{-3}$ to $1$
for $m_{\chi}\in[10^{-3},1]\ {\rm MeV}$. This is rather similar to the stable $Z'$ case but note that we are now studying the $2m_\chi\leq m_{Z'}$ part of the parameter space while the stable case requires $2m_\chi\geq m_{Z'}$ 
It is conceivable that the $Z'$ decays to other dark sector states that interact with the DM via couplings that are independent from the one that governs the $Z'$ decay. In this case significantly smaller $Z'$ couplings allow for a complete transfer of the energy to the dark sector. We refrain from further discussions on this possibility here since we want to work with a minimal set of new particles.

Within the mass range indicated by Fig.~\ref{fig:g-chi}, the correct relic abundance of DM can easily be achieved via the freeze-in mechanism and such DM candidates can also be easily accommodated in various complete models~\cite{Dvorkin:2019zdi}.
A well-known example in this mass range is keV sterile neutrino DM---see
Refs.~\cite{Dasgupta:2021ies,Abazajian:2017tcc,Drewes:2016upu} for
reviews. Although the simplest scenario of sterile neutrino DM has
been in tension with X-ray observations and bounds from structure
formation, new interactions of sterile neutrinos can easily revive
their feasibility as DM candidates---see, e.g., ~\cite{Astros:2023xhe,Hansen:2017rxr}.
The coupling $g_{\chi}$ may be subjected to some cosmological constraints,
depending on various factors including whether $\chi$ had been thermalized
in the early universe, whether it is asymmetric DM. For $\chi$ in
the sub-MeV regime, large $g_{\chi}$ is allowed by existing cosmological
constraints, provided that $g_{\psi}$ is suppressed such that $\sqrt{g_{\chi}g_{\psi}}\lesssim10^{-6}$~\cite{Hufnagel:2021pso,Wang:2023csv}.

\section{Benchmark models \label{sec:models}}

In this section we move to concrete particle physics models. We restrict ourselves to four simple representative benchmark cases: the dark photon, a $Z'$ from either gauged $U(1)_{B-L}$ or gauged $U(1)_{L_{\tau}-L_{\mu}}$, and the dark Higgs.
The first three of these are models with vector mediators. Despite this seeming similarity they differ strongly in the coupling structures which has a profound impact on the phenomenology. The last one features a light scalar that interacts with the SM through mixing with the Higgs. Clearly, these do not cover all possibilities but we believe that they give a reasonable cross section through the space of available models and illustrate nicely that we can fulfill the model-independent requirements for energy injection into the DM halo while respecting model-dependent constraints from other observables. 

In the following we we will go through the models one by one. For each of them we will briefly  introduce the model, comment on which production rates discussed in Sec.~\ref{sec:particle}  are needed in this case, compare the results with other constraints, and identify the regions of parameter space that allow a significant impact on the DM halo.

\begin{figure}
\centering

\includegraphics[width=0.49\textwidth]{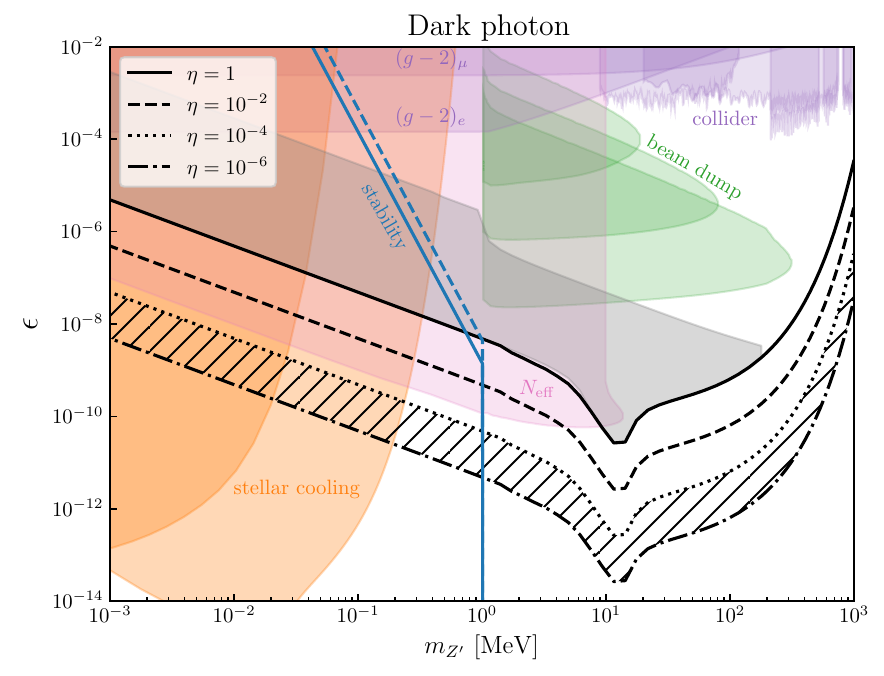}\includegraphics[width=0.49\textwidth]{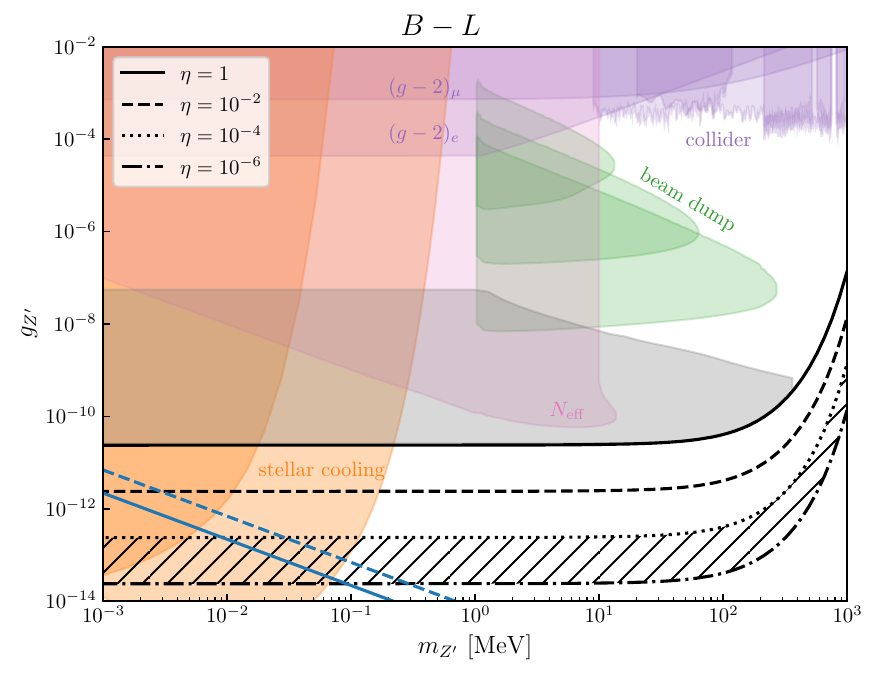}

\includegraphics[width=0.49\textwidth]{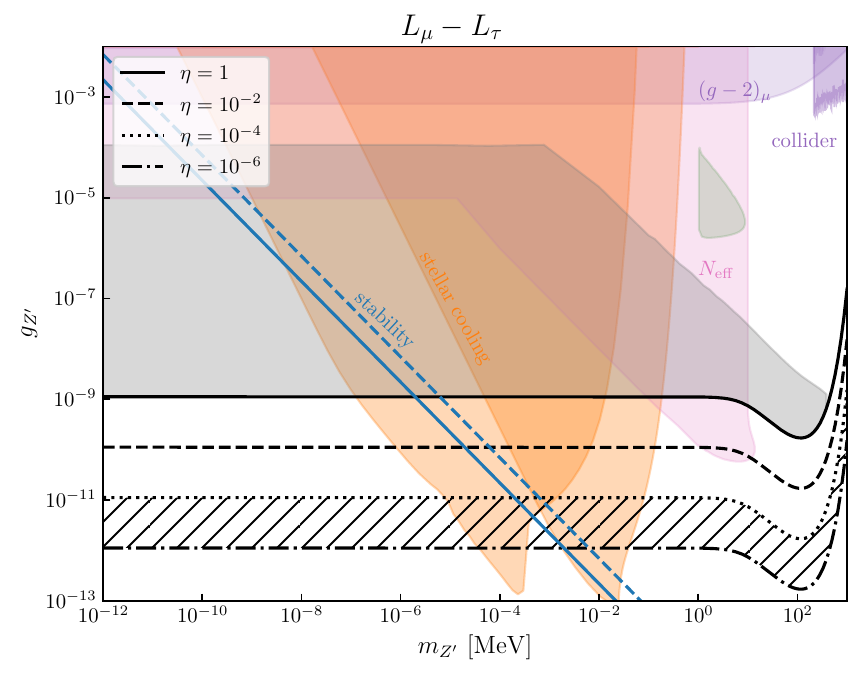}\includegraphics[width=0.49\textwidth]{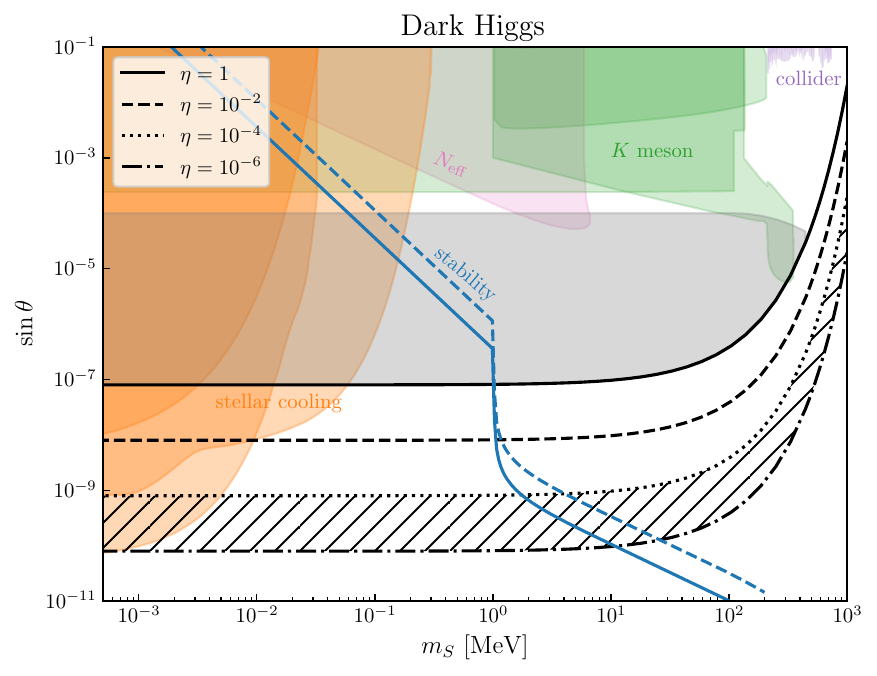}

\caption{
Parameter space for specific models. The black lines represent the
required  coupling strength of the dark radiation with the SM to
generate the desired luminosity of dark radiation---see Eq.~\eqref{eq:eta} for the definition of $\eta$ and Fig.~\ref{fig:exotics_fraction} for the astrophysical constraint on $\eta$. The gray regions represent the usual SN cooling bounds in the absence of dark sector couplings. The hatched regions correspond to the scenario that SNe  deposit the required amount of energy into the DM halo for the cusp-core transformation. 
The lower line of the hatched region corresponds roughly to the value needed to explain the observed cores base on our preferred profile  Eq.~\eqref{eq:rho_core} while the upper line approximates the upper limit based on the more conservative halo profile in Eq.~\eqref{eq:coredNFW}. 
The colored shaded regions are excluded by known bounds from stellar cooling, the cosmological
$N_{{\rm eff}}$, beam dump experiments, and collider searches---see
the text for further explanations.  The blue lines
indicate the stability of the dark radiation at relevant  astrophysical
scales: below the solid (dashed) blue lines the dark radiation can
travel more than $0.1$ kpc (1 kpc) before decay. 
Above the blue lines, we assume that $Z'$ or $S$ dominantly and instantly decays to DM. In this case, the usual trapping mechanism does not apply directly. Nevertheless, the produced DM could be trapped in the SN and the ``open'' region above the SN cooling region ought to be investigated in a dedicated analysis. 
The dominance of the dark decay mode also implies that the presented bounds from beam dump and collider searches, which typically look for the decay of $Z'$/$S$ into SM, may be significantly diminished.
\label{fig:space}}
\end{figure}

\subsection{Dark photon}

The dark photon model assumes the presence of a spontaneously broken dark $U(1)$ gauge
symmetry whose gauge boson interacts with the SM only via the kinetic
mixing portal~\cite{Holdom:1985ag}. More specifically, the SM hypercharge
gauge boson could be coupled to a massive dark $U(1)$ gauge boson via
\begin{equation}
{\cal L}\supset-\frac{\epsilon}{2}F^{\mu\nu}F'_{\mu\nu}\thinspace,\label{eq:-31}
\end{equation}
where  $F^{\mu\nu}$ and $F'_{\mu\nu}$ are the gauge field strength
tensors of the SM $U(1)_{Y}$ and the dark $U(1)$, respectively. 

Although Eq.~\eqref{eq:-31} implies couplings to both the SM $Z$ boson
and the photon, the dark gauge boson in the low-mass limit (well below
the $Z$ boson mass) behaves as a photon-like boson, i.e.~its effective
couplings to the SM fermions generated by the kinetic mixing are proportional
to their electric charges\,---\,see e.g.~discussions in Ref.~\cite{Li:2023vpv}.
This allows us to consider the simplified dark photon model that contains
only the kinetic mixing with the photon\footnote{Although it no longer respects the  gauge invariance of the SM, the
simplified dark photon model can be regarded as a low-energy effective
theory of the complete theory in Eq.~\eqref{eq:-31}. This is perfectly adequate in the regime of interest to us since the largest energies considered here are $\mathcal{O}(100)$ MeV.}, and we can assume $F_{\mu\nu}$ is the field strength tensor of
the photon, with $\epsilon$ replaced by 
\begin{equation}
\varepsilon\equiv\epsilon\cos\theta_{W}\thinspace,\label{eq:-32}
\end{equation}
where $\theta_{W}$ is the Weinberg angle.  In the physical basis
where both the photon and the dark photon are in mass eigenstates
and their kinetic terms have been canonically normalized, we denote the dark
photon by $Z'$ and its mass by $m_{Z'}$. In this basis, the canonicalization of the  kinetic terms gives rise to the following effective couplings of $Z'$ to SM fermions:
\begin{equation}
g_{\psi}=\varepsilon eQ_{\psi}\thinspace,\label{eq:-33}
\end{equation}
where $e=\sqrt{4\pi\alpha}$ and $Q_{\psi}$ is the electric charge
of $\psi$. 

Note that $g_{\psi}$ in Eq.~\eqref{eq:-33} is the
effective coupling in vacuum. 
In SN, the dark photon is produced in a dense and hot medium where mediums effect can be significant. 
The medium modifies
the photon self-energy, implying that the photon and the dark photon,
which are  mass eigenstates in vacuum,  are no longer mass eigenstates
in the medium. Adjusting the basis accordingly, the effective coupling is modified to
\begin{equation}
g_{\psi,m}\approx g_{\psi}\left|\frac{m_{Z'}^{2}}{m_{Z'}^{2}-\Pi_{\gamma\gamma}}\right|\thinspace,\label{eq:-34}
\end{equation}
where $\Pi_{\gamma\gamma}$ denotes the medium contribution to the
photon self-energy. Note that in the limit of $m_{Z'}\to0$, Eq.~\eqref{eq:-34}
vanishes, which implies the dark photon would be decoupled from the
plasma and cannot be effectively produced via thermal processes. This
is a unique feature of the dark photon (see Ref.~\cite{An:2013yfc}
for more discussions) and it requires that the vacuum coupling in
Eq.~\eqref{eq:-33} is proportional to the electric charge.

Due to its equal couplings to the electron and proton and the absence of couplings to neutrinos, the dark photon is produced dominantly via  NBr-2 and SC. 
The production rate of the dark photon can be straightforwardly computed
using the results in Sec.~\ref{subsec:Dominant-production}, with 
\begin{equation}
\alpha_{\psi}'=\alpha\varepsilon^{2} Q_{\psi}^2 \frac{m_{Z'}^{4}}{\left(m_{Z'}^{2}-{\rm Re}\Pi_{\gamma\gamma}\right)^{2}+\left({\rm Im}\Pi_{\gamma\gamma}\right)^{2}}\thinspace.\label{eq:-35}
\end{equation}

The total luminosity of $Z'$ is composed of
\begin{equation}
L_{Z'}=2L_{Z',T}+L_{Z',L}\thinspace,\label{eq:-36}
\end{equation}
where $L_{Z',T}$ and $L_{Z',L}$ denote the contributions of transverse
and longitudinal polarization modes, respectively. For each mode,
we use Eq.~\eqref{eq:-12} to compute the contribution.  The detailed
calculation involves proper handling of the real and imaginary parts
of $\Pi_{\gamma\gamma}$, which are also polarization dependent, as
well as a careful treatment of the resonance that occurs at $m_{Z'}^{2}\to{\rm Re}\Pi_{\gamma\gamma}$
in Eq.~\eqref{eq:-35}. The  details are explained in Appendix~\ref{sec:The-medium-effect}. 

In Fig.~\ref{fig:space}, the upper left panel, we plot four black
contours to indicate the required magnitude of the kinetic mixing
to generate $\eta=1$, $10^{-2}$, $10^{-4}$, and $10^{-6}$.
We comment here that due to $L_{Z',T}\propto\alpha\varepsilon^{2}m_{Z}^{4}$
and $L_{Z',L}\propto\alpha\varepsilon^{2}m_{Z}^{2}$ in the low-mass limit (see also Appendix~\ref{sec:The-medium-effect}), the production of very light dark photon is actually dominated by the longitudinal mode. As a consequence, the black curves have the asymptotic behavior of $\epsilon\propto 1/m_{Z'}$ in the low-mass limit. This is a unique feature of the dark
photon model.

To assess which parts of the parameter space are still open, we add bounds from existing experiments  in the relevant mass
range. 
Here the stellar cooling
bounds are taken from Ref.~\cite{Li:2023vpv}, derived from observations
of the Sun and red giants. Bounds from laboratory searches can be
readily produced via the {\tt DARKCAST} package~\cite{Ilten:2018crw}.
In this plot, the beam dump limits are produced by {\tt DARKCAST}
using data sets from E137~\cite{Bjorken:1988as}, E141~\cite{Riordan:1987aw}, and Orsay experiments~\cite{Davier:1989wz}; the collider
limits are produced using data sets from BaBar~\cite{Lees:2014xha}, NA48~\cite{Batley:2015lha}, and LHCb~\cite{Aaij:2019bvg} experiments;
and the $(g-2)_{\mu,e}$ bounds are derived from anomalous magnetic
moments of the muon and the electron. In addition, light $Z'$ around
or below the MeV scale could be thermalized in the early universe
and modify the cosmological effective number of neutrino species ($N_{{\rm eff}}$).
So we also impose a constraint from $N_{{\rm eff}}$ on this plot,
taken from Ref.~\cite{Li:2023puz}.

As previously discussed, the viable parameter space should be interpreted
differently for stable and unstable $Z'$. For the dark photon, which
does not couple to neutrinos, the dominant decay is $Z'\to2e$ if
$m_{Z'}$ is above $2m_{e}$. According to the estimate in Sec.~\ref{subsec:Lifetime-opacity},
the magnitude of $\epsilon$ for $Z'$ with $m_{Z'}>2m_{e}$ to be
stable at relevant astrophysical scales is lower than around $10^{-15}$,
well below the plot range of the plot for the dark photon in Fig.~\ref{fig:space}.
For $m_{Z'}\ll 2m_{e}$, the dominant decay channel is $Z'\to3\gamma$,
which has the following decay rate~\cite{Redondo:2008ec}\footnote{Eq.~\eqref{eq:-43} is derived from the Euler-Heisenberg limit which requires $m_{Z'}\ll 2m_{e}$.  For $m_{Z'}$ comparable to $m_{e}$ but less than  $2m_{e}$, the deviation from this limit can be significant---see e.g.~Refs.~\cite{McDermott:2017qcg,Linden:2024uph,Linden:2024fby}. According to Fig.~3 of Ref.~\cite{McDermott:2017qcg}, this deviation is less than $30\%$ for $m_{Z'}< 0.4$ MeV. Since the range with significant deviation is relatively narrow (only from 0.4 to 1 MeV), we do not include the correction to the Euler-Heisenberg limit in our analysis.}:
\begin{equation}
\Gamma_{Z'\to3\gamma}=\frac{17\alpha^{4}\varepsilon^{2}}{11664000\pi^{3}}\frac{m_{Z'}^{9}}{m_{e}^{8}}\thinspace.\label{eq:-43}
\end{equation}
Using Eq.~\eqref{eq:-43}, we plot two blue lines corresponding to
$l_{{\rm decay}}=1$ kpc (solid) and $0.1$ kpc (dashed) in the upper
left panel of Fig.~\ref{fig:space}. Below the blue lines, the dark
photon can be stable at relevant astrophysical scales. Above the blue
lines, we assume $m_{Z'}>2m_{\chi}$ such that it decays dominantly
to DM.

\subsection{Dark $Z'$ from $U(1)$ extensions}

There are a few possibilities to extend the SM gauge symmetry by an
extra $U(1)$ under which the SM fermions are charged and hence directly
interact with the gauge boson arising from the extra $U(1)$. By requiring
that the extra $U(1)$ is anomaly free, the most commonly considered
possibilities are $B-L$, $L_{e}-L_{\mu}$, $L_{e}-L_{\tau}$, and
$L_{\mu}-L_{\tau}$.   Due to phenomenological similarities between
$L_{e}-L_{\mu(\tau)}$ and $B-L$, we only select $B-L$ and $L_{\mu}-L_{\tau}$
for case studies. The couplings in these models to medium particles at tree-level are given by 
\begin{equation}
    g_{\psi}=g_{Z'}Q'_{\psi}\,, \label{eq:gZ}
\end{equation}
where $g_{Z'}$ is the gauge coupling of the extra $U(1)$ symmetry and  $Q'_{\psi}$ denotes the charge of $\psi$ under this symmetry. For the $B-L$ model, we have $Q'_{\psi}=1$ for all baryons and $Q'_{\psi}=-1$ for all leptons. For the $L_{\mu}-L_{\tau}$ model,  $Q'_{\psi}$ takes $1$ or $-1$ for $\mu$- or $\tau$-flavored leptons, respectively.  Although the electron and quarks are not directly coupled to the $Z'$ in the $L_{\mu}-L_{\tau}$ model, they can be indirectly coupled via a $\mu$ or $\tau$ loop. The loop-induced couplings are about a factor of $10^{-3}$ smaller than the direct coupling to $\mu$ and $\tau$---see e.g.~\cite{Coy:2021wfs}. Some bounds from beam dump and neutrino scattering experiments actually rely on the loop-induced couplings.

The production rates of $Z'$ in both models can be straightforwardly
computed by rescaling the curves in Fig.~\ref{fig:g_bound}. 
For the $B-L$ model, the dominant production channels are NBr-4, SC, and $\nu$Co.
As for the $L_{\mu}-L_{\tau}$ model, due to the absence of tree-level couplings to the electron
and quarks, the dominant production channels are SC-$\mu$ and $\nu$Co. The aforementioned loop-induced couplings have little impact on the SN production of $Z'$ in the $L_{\mu}-L_{\tau}$ model. Taking into account these dominant production channels, we obtain the black lines presented in the upper right and lower left panels in Fig.~\ref{fig:space}.  For both models, the stability curves (blue)
are determined by the decay width of $Z'\to2\nu$, which has been
computed in Eq.~\eqref{eq:-39}.

Regarding the existing bounds on these two models, we also run the {\tt DARKCAST} package and impose the obtained bounds on the plots. The beam dump and collider bounds on the $B-L$ gauge boson are similar to those on the dark photon, as can be expected from their similarities in couplings to the electron and the proton. As for the $L_{\mu}-L_{\tau}$ model, the loop-induced couplings are already included in the model file provided by {\tt DARKCAST} but they cause negligibly weak bounds in most cases. In particular, the beam dump constraint on the $Z'$ in this model is weak because its decay in the low-mass regime is dominated by the invisible mode $Z'\to2\nu$. The collider bounds on such a muonphilic $Z'$ are derived from searches 
for $4\mu$ final states ($e^{+}e^{-}\to\mu^{+}\mu^{-}Z'$ with $Z'\to\mu^{+}\mu^{-}$)~\cite{BaBar:2016sci,CMS:2018yxg}. 

In addition to laboratory bounds, there are also astrophysical and cosmological bounds. The cosmological $N_{\rm eff}$ bounds are taken from Ref.~\cite{Li:2023puz}, which computed the production of $Z'$ in the early universe via $\nu+\overline{\nu}\to Z'+Z'$ and $\nu+\overline{\nu}\to Z'$. The latter usually dominates over the former since the squared amplitudes of these two processes are proportional to $g_{\nu}^4$ and  $g_{\nu}^2$, respectively. However, in the ultralight regime of $Z'$, $\nu+\overline{\nu}\to Z'$ is suppressed by small $m_Z'$ while $\nu+\overline{\nu}\to Z'+Z'$ is not. Consequently, for the $L_{\mu}-L_{\tau}$ model which extends to the ultralight regime in Fig.~\ref{fig:space}, the  $N_{\rm eff}$ bound becomes flat at very small $m_{Z'}$. Note that here we have not taken into account its interactions with electrons, which would modify the $N_{\rm eff}$ bound on the $B-L$ gauge boson around the MeV scale~\cite{Esseili:2023ldf}. 
In the plots for  $B-L$ and $L_{\mu}-L_{\tau}$, we also impose stellar cooling bounds from Ref.~\cite{Li:2023vpv}. 
These bounds rely on electron and nucleon couplings, which in the $L_{\mu}-L_{\tau}$ are induced at the one-loop level, as we have mentioned above. These  loop-induced couplings are photon-like, similar to the dark photon case, causing a suppressed production rate of $Z'$ in the ultralight regime in ordinary  stellar medium. Hence the stellar cooling bound on $L_{\mu}-L_{\tau}$ vanishes in the limit of $m_{Z'}\to 0$. This is however not the case in the SN core, where abundant muons directly participate in the production of $Z'$.

Our analyses for these two models suggest that the $B-L$ model is viable only in the regime
of unstable $Z'$, while $L_{\mu}-L_{\tau}$ allows for both stable
and unstable regimes.

\subsection{Dark Higgs}

Another well-motivated particle to serve as the SN energy carrier
is the dark Higgs, which is a neutral scalar and interacts with the
SM via mass mixing with the SM Higgs.   Consequently,  its couplings
to SM fermions are proportional to the Higgs couplings to them, i.e.,
\begin{equation}
g_{\psi}=y_{\psi}\sin\theta\thinspace,\label{eq:-37}
\end{equation}
where $g_{\psi}$ and $y_{\psi}$ are the effective couplings of the dark Higgs 
and the SM Higgs with $\psi$, and $\theta$ denotes the mass mixing
angle.  Following the convention in the literature, we denote the
dark Higgs by $S$ and its mass by $m_{S}$. 

For fundamental fermions like $e$ and $\mu$, $y_{\psi}$ is determined
by the fermion masses: $y_{\psi}=\sqrt{2}m_{\psi}/v_{\text{EW}}$
with $v_{\text{EW}}\approx246$ GeV.  For nucleons, the effective
couplings are approximately the same for protons and neutrons: $y_{p}\approx y_{n}\approx2.2\times10^{-3}$~\cite{Cheng:1988im}\footnote{We note here that this value includes an important contribution from
the $s$ quark, rendering it significantly larger than the old value,
$y_{p}\approx y_{n}\approx8.5\times10^{-4}$ \cite{Shifman:1978zn},
which is used in some of the relevant studies---see e.g.~\cite{Hardy:2016kme,Yamamoto:2023zlu}. }. 

The production of the dark Higgs mainly relies on $g_{n,p}$, $g_{e}$,
and $g_{\mu}$. Due to the significant muon abundance in SN
and $g_{\mu}\gg g_{e}$, we find that muons actually leads to a larger
contribution to the production than electrons. Nevertheless, we include
the contributions of both muons and electrons as well as the dominant
one from NBr. The result is presented in the lower right panel of
Fig.~\ref{fig:space}.

In this plot, the stellar cooling bounds are taken from Ref.~\cite{Yamamoto:2023zlu},
derived from white dwarfs, red giants, and horizontal branch stars.
Due to the relatively small coupling of the dark Higgs to the electron,
the constraints from beam dump experiments are typically very weak.
Instead, measurements of the $K$ meson decay set stronger constraints
on the dark Higgs. Here we take the $K$ meson bounds from \cite{Dev:2017dui}
 and impose them on the plot, together with collider bounds obtained
from {\tt DARKCAST}.

The decay width of the dark Higgs to SM finals states can be found easily by taking the results for a SM Higgs from e.g. \cite{Djouadi:2005gi},  replacing  $m_h$ with $m_S$, and rescaling the coupling with $\sin \theta $. 
For $m_S$ above $2 m_e$ but below $2 m_\mu$,  
the width is dominated by the decay into electron positron pairs:
\begin{equation}
\Gamma_{S\to e^+ e^-}=\frac{m_{S}m_{e}^{2}\sin^{2}\theta}{8\pi v_{\text{EW}}^{2}}\left(1-\frac{4m_{e}^{2}}{m_{S}^{2}}\right)^{3/2}\\,\label{eq:-45}
\end{equation}
where $v_{\text{EW}}$ is the vacuum expectation value of the Higgs field.

The decay channel of the dark Higgs relevant to our analysis for $m_{S}<2m_{e}$ is
$S\to2\gamma$.
It involves triangle loop diagrams with SM charged particles
(quarks, charged leptons, and the $W^{\pm}$ boson).
Using the same prescription as before one finds for $m_S \ll M_W$
the partial decay widths
\begin{equation}
\Gamma_{S\to2\gamma}\approx\frac{m_{S}^{3}}{8\pi v_{\text{EW}}^{2}}\cdot \sin^2 \theta \cdot\frac{\alpha^{2}}{18\pi^{2}}\cdot\left|\sum_{\psi}Q_{f}^{2}N_{c,\psi}-\frac{21}{4}\right|^{2},\label{eq:-21}
\end{equation}
where 
$Q_{\psi}$ denotes the electric charge of the SM fermion $\psi$,
and $N_{c,\psi}=3$ or $1$ for quarks or leptons, respectively.

Using these equations
we compute the lifetime of the dark Higgs for $m_{S}<2m_{\mu}$ and
add the corresponding blue lines in the lower right panel of Fig.~\ref{fig:space}. Note that the width above the electron positron threshold is still suppressed by the small electron Yukawa compared to the $Z'$ benchmark models. Therefore, long lifetimes are possible even for $m_S> 1$ MeV and we find some region of the parameter space where the dark radiation is stable on halo scales at such high masses.

\section{Conclusions \label{sec:conclusion}}

Type II SN explosions release large amounts of energy over the lifetime of a galaxy. While the bulk of this energy is expected to go into neutrinos in the Standard Model, the observational limits on the amount of energy lost to a dark sector are weak. Up to an order one fraction of the energy released in the explosion can go to exotic light degrees of freedom. 
In this work, we investigated the possibility that this energy is not lost but absorbed by the DM halo. While the total amount of energy that can be made available from SN explosions is small compared to the overall binding energy of the DM halo, it is large enough to have an appreciable effect on the structure of the halo. Interestingly, observations of some dwarf galaxies are at variance with the expectation from DM-only simulations in that they prefer a cored halo while simulations point towards cuspy ones. Using an ansatz for a cored profile that recovers the unperturbed NFW one at large distances, we compute the energy needed to create a DM core of a certain size in an originally cuspy halo.

 We consider a set of classical dSphs analyzed in \cite{Read:2018fxs}. Observations show a mixed picture with some galaxies preferring a core while others are consistent with an NFW profile. Out of the eight dSphs considered here, two show a preference for a core at $2\ \sigma$ C.L. and six at $1\ \sigma$ C.L. while two only permit to place an upper limit on the size of a core. Taking the $2\ \sigma$ upper limit on the core radius we derive an upper limit on the amount of energy that can be absorbed by the DM halo. This can be interpreted as an upper limit on the fraction of energy released into the dark sector. In addition, two dSphs show a clear $2\ \sigma$ preference for a core while a further 6 show at least a mild (1-2 $\sigma$) preference. Interestingly, the preferred core size in all these systems points towards a rather similar fractional energy release from SN explosion in the ballpark of a few times $10^{-6}$. It would be interesting to investigate further to what extent this can explain the cusp vs core or the diversity problem of dwarf galaxies.

The above argument is relatively general and does not rely on a particular particle physics model. Nevertheless, it is a very important question if a model that fulfills the basic requirement, i.e.~appreciable production of light particles in SN explosion and subsequent energy transfer to the DM halo, exists and which parts of the parameter space support the mechanism for coring the DM halo. 
To answer this question, we first provide some general results for the emission of a general light vector boson ($Z'$) serving as dark radiation from SN explosions and its mean free path in the DM halo. We identify qualitatively different situations that can be classified according to the mass hierarchy between the DM and the particle emitted in the explosion. On the one hand, for $2 m_\chi< m_Z'$, the produced particles decay to DM on length scales that are short compared to the DM halo. In this case, energy transfer proceeds via elastic scattering between the energetic DM particles produced in the explosion and the non-relativistic particles that make up the halo.
On the other hand, for $2 m_\chi> m_{Z'}$, the $Z'$ particle can be stable if the couplings to the SM are small enough. In this case the energy transfer proceeds via Compton scattering of $Z'$ on $\chi$. Both cases point towards light DM candidates with masses of up to $10$ MeV and relatively large couplings between the DM and $Z'$. 

Equipped with these general estimates we studied four representative benchmark models: the dark photon, a light $Z'$ from spontaneously broken $U(1)_{B-L}$ or $U(1)_{L_\mu -L_\tau}$, and the dark Higgs. In all these models we find some range of parameters that can lead to an energy deposit in the DM halo in excess of what is allowed based on the observed upper limits on the core radius. Therefore, we conclude that further studies of the DM halo can open the way to new tests of BSM physics and can extend the reach of the limits based on SN observations considerably.

\acknowledgements

SV acknowledges support by the DFG via individual research grant Nr. 496940663. SV also thanks the Mainz Institute for Theoretical Physics (MITP) for hospitality during the workshop ``The Dark Matter Landscape: From Feeble to Strong Interactions", where part of this work was performed.  The work of X.\,J.\,X is supported in part by the National Natural Science Foundation of China (NSFC) under grant No.~12141501 and also by the CAS Project for Young Scientists in Basic Research (YSBR-099). X.\,J.\,X would also like to thank the Peng Huanwu Center for Fundamental Theory (PCFT) in Hefei for the hospitality and financial support (NSFC grant No.~12247103) during his visit when part of this work was performed.

\appendix

\section{The medium effect \label{sec:The-medium-effect}}

In this appendix, we briefly review some formulae in plasma physics
used in our work, and discuss the medium effect which is particularly
important to the dark photon model. 

In the finite temperature field theory, the production and absorption
rates (also referred to as the gain and loss rates and hence denoted
by $\Gamma_{\gamma}^{(\text{gain})}$ and $\Gamma_{\gamma}^{(\text{loss})}$
below) of the photon are related to the imaginary part of the photon
self-energy (${\rm Im}\Pi_{\gamma\gamma}$) in the medium by~\cite{Weldon:1983jn}
\begin{equation}
\Gamma_{\gamma}^{(\text{gain})}=f_{\gamma}\Gamma_{\gamma}\thinspace,\ \ \Gamma_{\gamma}^{(\text{loss})}=(1+f_{\gamma})\Gamma_{\gamma}\thinspace,\ \ \Gamma_{\gamma}=-\omega^{-1}{\rm Im}\Pi_{\gamma\gamma}\thinspace,\label{eq:-46}
\end{equation}
where $f_{\gamma}=\left(e^{\omega/T}-1\right)^{-1}$ with $\omega$
the photon energy.  Since $\Gamma_{\gamma}^{(\text{loss})}$ is easier
to compute than $\Gamma_{\gamma}^{(\text{gain})}$, the former is
often used to determined the latter via
\begin{equation}
\Gamma_{\gamma}^{(\text{gain})}=\frac{f_{\gamma}}{1+f_{\gamma}}\Gamma_{\gamma}^{(\text{loss})}=e^{-\omega/T}\Gamma_{\gamma}^{(\text{loss})}\thinspace.\label{eq:gain-loss}
\end{equation}
Note that the above relations are only valid for the photon which
is in equilibrium. For  $Z'$ considered in this work, which is 
not in equilibrium, we  can approximately estimate its production
rate from $\Gamma_{\gamma}^{(\text{gain})}$ with proper substitution
of relevant couplings. 

The real part of the photon self-energy is given by
\begin{equation}
{\rm Re}\Pi_{\gamma\gamma}=\begin{cases}
\omega_{P}^{2} & \text{for}\ T\ \text{polarization}\\
\omega_{P}^{2}\left(1-\frac{|\mathbf{k}|^{2}}{\omega^{2}}\right) & \text{for}\ L\ \text{polarization}
\end{cases}\thinspace,\label{eq:-20}
\end{equation}
where $\mathbf{k}$ is the momentum of the photon and $\omega_{P}$
is the plasmon frequency. In the SN medium with high electron degeneracy,
one can use the following formula for the plasmon frequency~\cite{Chang:2016ntp}:
\begin{equation}
\omega_{P}^{2}=\frac{4\pi\alpha n_{e}}{E_{F}}\thinspace,\label{eq:-7}
\end{equation}
with $E_{F}$ the fermi energy of electrons:
\begin{equation}
E_{F}\equiv\sqrt{m_{e}^{2}+\left(3\pi^{2}n_{e}\right)^{2/3}}\thinspace.\label{eq:-5}
\end{equation}

As mentioned in the main text, the medium effect may lead to effective
couplings of $Z'$ that are very different from the vacuum ones:
\begin{equation}
g_{\psi}\xrightarrow{\text{medium}}g_{\psi,m}\thinspace.\label{eq:-1}
\end{equation}
This change is caused by the in-medium mixing between $Z'$ and the
photon, which is essentially an effect of coherent scattering of $\gamma+\psi\leftrightarrow Z'+\psi$\,---\,see
Appendix B of Ref.~\cite{Li:2023vpv}. Consequently, the medium effect
crucially depends on how $Z'$ is coupled to charged particles in
the medium.  Assuming that the charged particles in the medium are
mainly electrons and protons, the medium effect modifies $g_{\psi}$
to~\cite{Li:2023vpv}
\begin{equation}
g_{\psi,m}^{2}=g_{\psi}^{2}\left|\frac{m_{Z'}^{2}-\kappa_{p}\Pi_{\gamma-(p)-\gamma}}{m_{Z'}^{2}-\Pi_{\gamma\gamma}}\right|^{2},\label{eq:-2-1}
\end{equation}
where $\Pi_{\gamma-(p)-\gamma}$ denotes the photon self-energy generated
by a proton in the loop, and $\kappa_{p}$ is defined to quantify
the deviation of $Z'$ couplings to photon-like couplings:
\begin{equation}
g_{p}:g_{e}=\kappa_{p}-1:1\thinspace.\label{eq:-3}
\end{equation}
Eqs.~\eqref{eq:-2-1} and \eqref{eq:-3} imply that if the $Z'$ couplings
are photon-like ($g_{p}:g_{e}=-1:1$), the effective coupling would
vanish in the $m_{Z'}^{2}\to0$ limit. Taking the dark photon model
for example, the effective kinetic mixing parameter ($\epsilon_{m}$)
is related to the vacuum one ($\epsilon$) by 
\begin{equation}
\epsilon_{m}^{2}=\epsilon^{2}\frac{m_{Z'}^{4}}{\left(m_{Z'}^{2}-{\rm Re}\Pi_{\gamma\gamma}\right)^{2}+\left({\rm Im}\Pi_{\gamma\gamma}\right)^{2}}\thinspace,\label{eq:-2}
\end{equation}
which agrees with Eq.~(1.2) in \cite{Chang:2016ntp} and has exactly
the vanishing feature at $m_{Z'}^{2}\to0$. 

\begin{figure}
\centering

\includegraphics[width=0.7\textwidth]{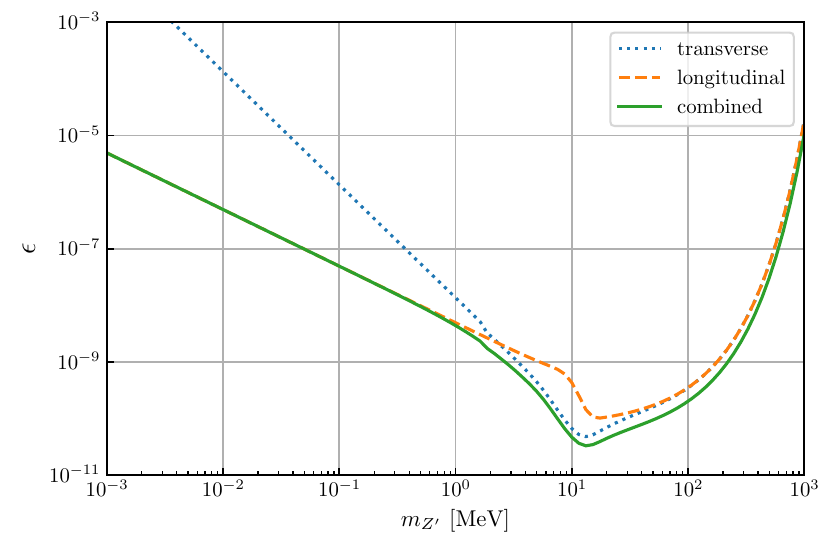}

\caption{
The required magnitude of the dark photon kinetic mixing to generate
$L_{Z'}=3\times 10^{52} {\rm erg}/{\rm s}$.
The dotted (dashed) line takes only one transverse
(longitudinal) production mode into account. The combined result (solid
line) is produced by adding contributions of two transverse and one
longitudinal modes together. \label{fig:dark-photon}}
\end{figure}

According to Eq.~\eqref{eq:-2}, the production rate of the dark photon
reads:
\begin{equation}
\Gamma_{Z'}^{(\text{gain})}=\epsilon_{m}^{2}\Gamma_{\gamma}^{(\text{gain})}=\epsilon^{2}f_{\gamma}\frac{m_{Z'}^{4}\Gamma_{\gamma}}{\left(m_{Z'}^{2}-{\rm Re}\Pi_{\gamma\gamma}\right)^{2}+\left(\omega\Gamma_{\gamma}\right)^{2}}\thinspace.\label{eq:-47}
\end{equation}
Note that Eq.~\eqref{eq:-47} implies a resonance at $m_{Z'}^{2}={\rm Re}\Pi_{\gamma\gamma}$.
When performing the integration in Eq.~\eqref{eq:-12} with the production
rate given above, this resonance can always be reached in the low-mass
regime. For the longitudinal polarization, this would cause a sharp
peak in $\int\text{d}k$ integral; for the transverse polarization,
it implies a peak in the $\int\text{d}r$ integral.  In practice,
these peaks  often cause numerical instability.   To overcome the
numerical instability, we adopt the delta-function approximation when
the resonance occurs.  This approximation makes use of the following
limit
\begin{equation}
\lim_{\Gamma\to0}\frac{\Gamma}{x^{2}+\Gamma^{2}}=\pi\delta(x)\thinspace,\label{eq:-48}
\end{equation}
which implies at the resonance Eq.~\eqref{eq:-47} can be viewed as
a delta-function. The specific value of $\Gamma_{\gamma}$ becomes unimportant at the resonance but it is still important to the production of the dark photon in the non-resonant zone with $\omega_P<m_{Z'}$.

With the above details being noted, it is straightforward to substitute
Eq.~\eqref{eq:-47} into Eq.~\eqref{eq:-12} and perform the integration
to obtain $L_{Z'}$. 
In Fig.~\ref{fig:dark-photon}, we show the
required magnitude of $\epsilon$ to generate $L_{Z'}=L_{\nu}$.
As  is shown in Fig.~\ref{fig:dark-photon}, the difference between
longitudinal and transverse production rates is very significant in
the low-mass regime. This can be understood from Eq.~\eqref{eq:-47}
where ${\rm Re}\Pi_{\gamma\gamma}$ and $\Gamma_{\gamma}$ in the
longitudinal mode contain an additional factor of $1-|\mathbf{k}|^{2}/\omega^{2}=m_{Z'}^{2}/\omega^{2}$
compared to those in the transverse mode. Consequently, the low-mass
limit becomes
\begin{equation}
\lim_{m_{Z'}\to0}\Gamma_{Z'}^{(\text{gain})}\propto\begin{cases}
\left(\epsilon m_{Z'}^{2}\right)^{2} & \text{for}\ T\ \text{polarization}\\
\left(\epsilon m_{Z'}\right)^{2} & \text{for}\ L\ \text{polarization}
\end{cases}\thinspace.\label{eq:-49}
\end{equation}
Therefore, in the low-mass regime, the dark photon production rate
should be dominated by the longitudinal emission\,---\,see Ref.~\cite{An:2013yfc}
for a more dedicated discussion.

\bibliographystyle{utphys.bst}
\bibliography{references.bib}

\end{document}